\documentclass[prb,aps,showpacs,twocolumn,preprintnumbers,amsmath,amssymb,superscriptaddress,longbibliography]{revtex4-1}
\usepackage[english]{babel}
\usepackage{amsmath,amssymb,amsfonts}
\usepackage{graphicx}
\usepackage[colorlinks=True,linkcolor=red,citecolor=blue,urlcolor=blue]{hyperref}
\usepackage{bookmark}
\usepackage{xcolor}
\usepackage{chngcntr}
\usepackage{braket}
\usepackage{nicefrac}
\usepackage{bm}
\usepackage{bbm}
\usepackage{braket}
\usepackage{array}
\newcolumntype{C}{>{$}c<{$}}

\newcommand{\diff}{\mathrm{d}}
\def\br{\mathbf{r}}
\def\bk{\mathbf{k}}
\def\id{\mathbbm{1}}

\def\ket#1{|#1\rangle }
\def\bra#1{\langle #1 |}

\begin{document}
\title{Topological Weaire-Thorpe models of amorphous matter}

\author{Quentin Marsal}
\affiliation{Univ. Grenoble Alpes, CNRS, Grenoble INP, Institut N\'eel, 38000 Grenoble, France}
\author{D{\'a}niel Varjas}
\affiliation{QuTech and Kavli Institute of Nanoscience, Delft University of Technology, 2600 GA Delft, The Netherlands}
\author{Adolfo G. Grushin}
\affiliation{Univ. Grenoble Alpes, CNRS, Grenoble INP, Institut N\'eel, 38000 Grenoble, France}

\date{\today}
\begin{abstract}
Amorphous solids remain outside of the classification and systematic discovery of new topological materials, partially due to the lack of realistic models that are analytically tractable. Here we introduce the topological Weaire-Thorpe class of models, which are defined on amorphous lattices with fixed coordination number, a realistic feature of covalently bonded amorphous solids. Their short-range properties allow us to analytically predict spectral gaps. Their symmetry under permutation of orbitals allows us to analytically compute topological phase diagrams, which determine quantized observables like circular dichroism, by introducing symmetry indicators for the first time in amorphous systems. These models and our procedures to define invariants are generalizable to higher coordination number and dimensions, opening a route towards a complete classification of amorphous topological states in real space using quasilocal properties. 
\end{abstract}
\maketitle

\noindent\textbf{Introduction}\\
Although most solids can be grown amorphous, their lack of translational symmetries has kept amorphous solids outside the recently developed topological classifications of non-interacting matter~\cite{Vergniory:2019ub,Zhang:2019tp,Tang18}, halting their discovery for robust applications. Amorphous Bi$_2$Se$_3$ was shown to be the sole exception recently, with spectral, spin and transport data supporting a surface Dirac cone\cite{Corbae:2019tg}. 
Other condensed matter platforms based on non-stoichiometric growth of the same compound are promising alternatives\cite{DC:2018uf,Sahu:wy} and, 
as a proof of principle, amorphous topological states have been realized in two-dimensional systems of coupled gyroscopes~\cite{Mitchell2018}.
However, the challenge is to model realistic materials, and determine their topological phase diagram in a way that may establish a classification, and aid their systematic discovery. 

Addressing this challenge seems possible since the absence of amorphous topological solids is not fundamental; topological protection does not rely on translational invariance.
{ This well developed understanding dates back at least to studies of integer quantum Hall transitions~\cite{Chalker_1988,Wei88,Huckestein95}.} 
{More recently, several classes of amorphous models have been shown to host integer quantum Hall (or Chern insulator) phases, as well as other topological states~\cite{Agarwala:2017jv,Xiao17,Poyhonen2017,Bourne:2018jr,Mitchell2018,agarwala2019higher,Yang19,Costa:2019kc,Mukati20,Sahlberg:2019uo}, including numerical work that suggests differences compared to known quantum Hall transitions~\cite{Sahlberg:2019uo,Ivaki2020}.}
Although {the corresponding topological phase diagrams} can be computed numerically, by simulating responses to external fields~\cite{Mukati20} or through real space topological markers~\cite{Agarwala:2017jv,Mitchell2018}, these methods are not generalizable to every discrete symmetry in every dimensionality.
{Crucially, a symmetry based approach~\cite{Kruthoff17,Po:2017ci,Bradlyn2017,Song:2018cj} for amorphous solids, which proved to be successful in high-throughput classifications of topological crystals~\cite{Vergniory:2019ub,Zhang:2019tp,Tang18}}, seems out of reach due to the absence of long-range atomic order.

In this work we find that an overlooked yet common property of covalently bonded amorphous solids, their fixed coordination number~\cite{Zallen}, can be exploited to overcome these problems. 
This property is rooted in the fact that the local chemical environment in an amorphous solid is similar to that of the crystalline phase of the same compound~\cite{Weaire:1971in,Toh:2020dy,Corbae:2019tg}. 
The local environment determines the coarse properties of the density of states such as spectral gaps, while long-range correlations or periodicity, determine the finer details.
Such physical input has been a cornerstone in describing amorphous states~\cite{Weaire71,Zallen}, allowing to prove the presence of spectral gaps in amorphous Si, eventually explaining why windows are transparent~\cite{Weaire:1971in,Weaire71,Thorpe:1973cr}.
{ Although topological properties are non-local in general, and quasi-local\cite{LCM} at best}, this useful chemical input remains unexploited in current models of amorphous topological states.

The models we propose are an analytically tractable and generalizable set of topological amorphous models with fixed coordination. 
They generalize the Weaire-Thorpe Hamiltonian class~\cite{Weaire71} explicitly developed to respect the local environment across sites.
We show analytically that they are generically gapped, and track the band crossings as a function of the parameters of the models.
Remarkably, these models allow us to construct an amorphous version of symmetry indicators by exploiting the symmetry resulting from the equivalence between orbitals.
We are able to map their topological phase diagram modulo an integer, without the need to compute local topological markers. 

\begin{figure}
    \centering
    \includegraphics[width=\columnwidth]{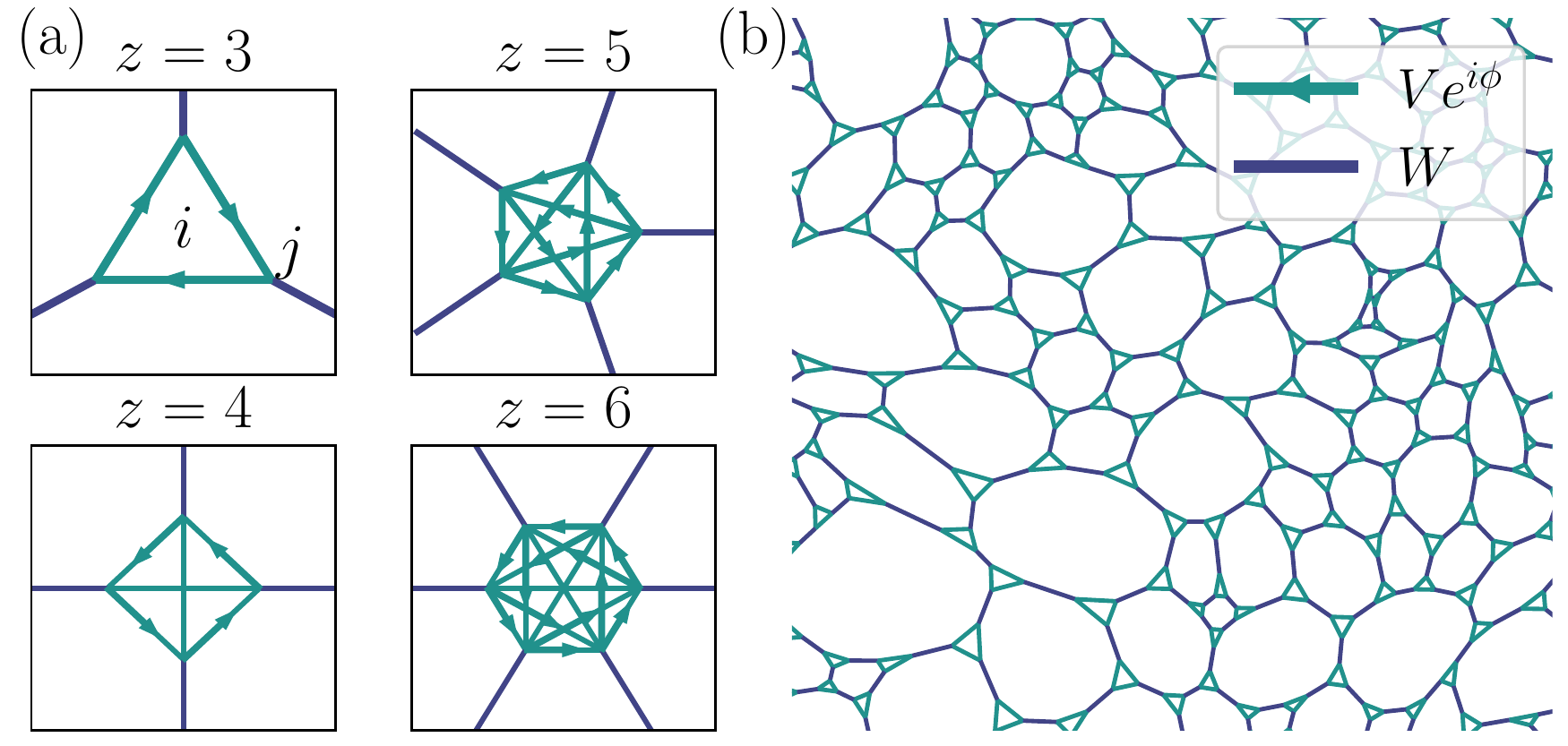}
    \caption{\textbf{Topological Weaire-Thorpe models.} (a) Basic building blocks for three-, four-, five-, six-fold coordinated Weaire-Thorpe models. The diagrams show a single site $i$ with $z$ orbitals labeled by $j$. The green lines indicate the intra-site hopping $Ve^{i\phi}$ while the blue lines are the inter-site hopping $W$. (b) Example of a three-fold coordinated topological Weaire-Thorpe model.}
    \label{fig:models}
\end{figure}

For concreteness, {in the main text} we exemplify our results using a triply coordinated two-dimensional amorphous lattice without time-reversal symmetry~\cite{Mitchell2018}, { and consider a fourfold coordinated model in the Supp. Mat.\cite{SuppMat} that emphasizes the generality of our results.} We analytically compute the spectral gaps and { numerically calculate} the ingap local density of states that shows topologically protected edge states. We numerically compute the local Chern marker~\cite{LCM}, which we link to a quantized circular dichroism, mapping the topological phase diagram in parameter space. We then introduce the symmetry indicators for this model and combine them in a formula that delivers the Chern number modulo three {(modulo four in the case of fourfold coordination)}, reproducing the topological phase diagram analytically. Finally{, for the threefold coordinated model, we discuss to which extent an} effective Hamiltonian approach\cite{Varjas2019}, that projects the full Hamiltonian into a basis of plane waves, can also detect topological phase transitions.

Due to their gapped structure and previous success in describing amorphous solids, the models we propose are natural candidates to describe realistic amorphous topological insulators and to track their topological phase transitions. 
Moreover, there is no fundamental restriction to extend our analytical arguments to different dimensionality, symmetry classes and coordination number, hinting at a route to classify amorphous topological insulators.

\section*{Results}

\noindent\textbf{Topological Weaire-Thorpe model class}\\
Irrespective of dimensionality and coordination, we define the topological Weaire-Thorpe models by a Hamiltonian with two terms
$H_{\mathrm{WT}} = H_V+H_W$, defined by
\begin{equation}
    H_\mathrm{WT} = \sum^{z}_{i,j\neq j'} V^{(i)}_{jj'}\ket{i,j}\bra{i,j'} + \sum^{z}_{i\neq i',j} W^{(j)}_{ii'}\ket{i,j}\bra{i',j}.
    \label{eq:Topo_WT}
\end{equation}
The index $i$ labels sites within a $z$ coordinated lattice, while $j=1,2,...z$ labels the $z$ orbitals within a site (see Fig.\ref{fig:models}(a)). The matrices $V^{(i)}_{jj'}$ connect different orbitals within a single site, while the matrices $ W^{(j)}_{ii'}$ connect different sites through a single orbital such that coordination remains fixed. If the intra and inter sites matrices are chosen real and independent of $i$ and $j$ respectively, such that $V^{(i)}_{jj'}=V\in \mathrm{Re}$ $\forall i,j\neq j'$ and $W^{(j)}_{ii'}=W \in \mathrm{Re}$ $\forall i\neq i',j$, {Eq.\eqref{eq:Topo_WT}} reduces to the Weaire-Thorpe model, introduced to describe spectral properties of tetravalent ($z=4$) amorphous materials such as amorphous Si~\cite{Weaire71}. The form of Eq.~\eqref{eq:Topo_WT} is motivated by the experimental observation that covalently bonded amorphous materials conserve the local environment imposed by their individual components, resembling their crystalline counterparts at short scales~\cite{Zallen}. Lattice disorder emerges at larger scales, modifying the lattice structure compared to the crystal (see  Fig.~\ref{fig:models}(b)).

To define the topological Weaire-Thorpe model here we allow $V$ to be complex, respecting that the local environment of different orbitals remains equivalent. This imposes that $V$ should be invariant with respect to cyclic permutation of the orbitals, and that the hopping between sites is fixed to $W$, { which we keep real}. These requirements do not fix the orientation of the { complex phases of $V$}, a freedom that can be adjusted depending on the physical context we wish to describe (see Fig.~\ref{fig:models}(a) for a specific convention and the Supp. Mat.\cite{SuppMat} { section \ref{AppendixB}} for further discussion). Given a convention for { these complex phases}, the fixed coordination will allow us to show that { Weaire-Thorpe models} have spectral gaps in general, and determine their band edges analytically.  

To show the existence of spectral gaps of the topological Weaire-Thorpe model and determine where they occur we use the resolvent method~\cite{Schwartz72}, outlined next and described in detail in the Supp. Mat.\cite{SuppMat} { section \ref{AppendixC}}. It is based on the observation that the eigenvalues of the system are poles of the complex function
\begin{equation}
    \label{eq:resolvent}
    \varepsilon \mapsto \frac{1}{\varepsilon-H_V-H_W} = \frac{1}{\varepsilon-H_V}\sum_{n=0}^{+\infty}\left(\frac{H_W}{\varepsilon-H_V}\right)^n,
\end{equation}
known as the resolvent of the Hamiltonian. If the series converges for a given $\varepsilon$, then $\varepsilon$ is not a pole of the resolvent and therefore not an eigenvalue of the Hamiltonian. A sufficient condition for convergence is that
\begin{equation}
    \left\lVert \frac{H_W}{\varepsilon-H_V} \right\rVert < 1,
    \label{eq:conv}
\end{equation}
where $\left\lVert\cdot\right\rVert$ is the operator norm, equal to the maximum absolute eigenvalue. Hence, we can determine the energy windows where there are no states, the spectral gaps, as a function of the model parameters with the condition $\min_{\lambda \in Sp(H_V)}\vert E-\lambda\vert>W$ where $\lambda$ spans the $z$ eigenvalues of $H_V$, $Sp(H_V)$. Although true for all $V$ and $W$, the criterion Eq.~\eqref{eq:conv} is useful as long as $W$ is less than the distance separating two eigenvalues of $H_V$, but it is not very informative when $W\gg V$. In this latter case, it is more instructive to use the freedom to interchange the roles of $H_V$ and $H_W$ in the last step of Eq.~\eqref{eq:resolvent}. In this case we arrive to the second condition $\min (E\pm W-\overline{V})>\max_{\lambda \in Sp(H_V)}(\vert\lambda-\overline{V}\vert)$, where $\overline{V}$ is a real number introduced to minimize $\Vert H_V-\overline{V}\Vert$.
The combination of these inequalities constrains the energy regions where there are no states, the spectral gaps. For generic $W$, $V$ and coordination $z$ these gaps are finite, and therefore the topological Weaire-Thorpe models describe an insulator at fillings where the chemical potential lies within the spectral gap. The gap boundaries are determined analytically by these inequalities, a useful property that we will use to determine the topological phase diagram.\\

\begin{figure}
    \centering
    \includegraphics[width=\columnwidth]{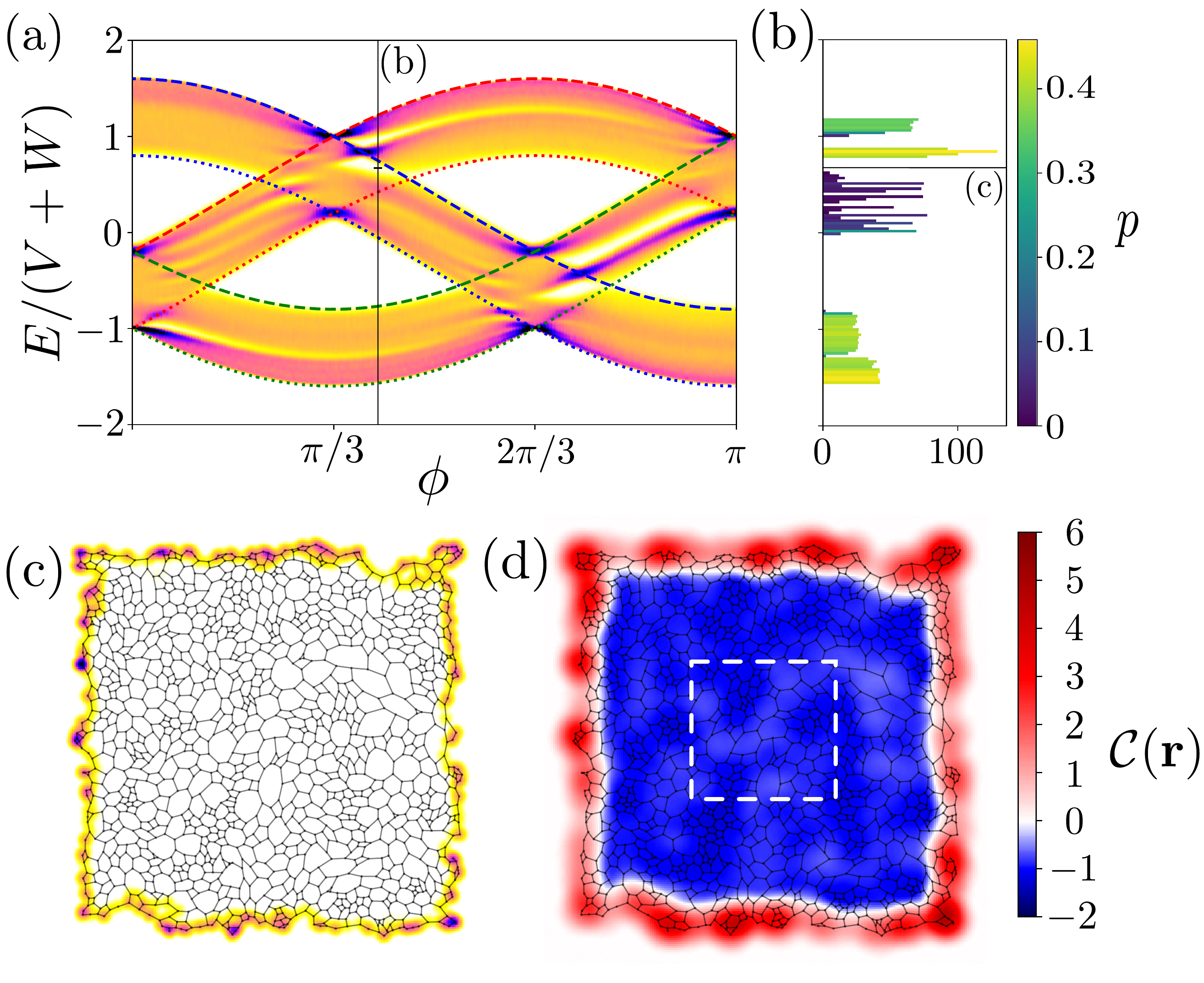}
    \caption{\textbf{Spectral properties} 
    (a) The energy spectrum for $W/V =0.66$ as a function of $\phi$ with periodic boundary conditions. The color intensity is proportional to the density of states (DOS). The dashed and dotted lines are obtained from the resolvent inequalities Eq.~\eqref{eq:3fbounds}, and correspond to states with $F_{+} = 1$, $F_{-} = 1$ respectively (see Fig.~\ref{fig:symmetry}(d)). Their color coding follows that of Fig.~\ref{fig:symmetry}(c). The vertical line $\phi=1.3$ indicates the parameters chosen for (b), (c), and (d). (b) The color bar shows the participation ratio $p= (\sum_{i}|\psi_i|^2)^2/N|\psi_{i}|^4$ for periodic boundary conditions, a measure of localization that indicates the ratio of sites contributing to the density of states within a given energy bin. The height of the histogram is proportional to the density of states. (c) The local density of states with open boundary conditions for an ingap state at 2/3 filling indicated by the black line in (b), showing the edge support of ingap states. (d) The local Chern marker density $c(\mathbf{r})$ at 2/3 filling, quantized to $C=-1$ in the bulk, with a large and positive edge contribution, typical of a Chern insulating phase. The white dashed square shows the averaging region used to compute Fig.~\ref{fig:symmetry}(a).}
    \label{fig:pd}
\end{figure}
\noindent\textbf{Three-fold coordinated Weaire-Thorpe-Chern insulator}\\
As an illustration of the power of { the Weaire-Thorpe models} we now construct a two-dimensional Hall insulator in an amorphous lattice with coordination $z=3$ and determine its electromagnetic responses and topological phase diagram numerically and analytically {(see Supp. Mat.\cite{SuppMat} section \ref{AppendixF} for the case $z=4$)}. With the building block labeled $z=3$ in Fig.~\ref{fig:models}(a) we first build a threefold coordinated lattice by making use of a Voronization procedure (see Supp. Mat.\cite{SuppMat} { section \ref{AppendixD}}). A specific realization of this lattice is shown in Fig.~\ref{fig:models}(b). As indicated by Fig.~\ref{fig:models}(a) we keep $W \in \mathrm{Re}$ and promote $V \to Ve^{i\phi}$ choosing the phases to connect in a clockwise fashion. 

As anticipated, the structure of the Weaire-Thorpe model allows to predict band gaps analytically. Applying our general criteria above we can determine the band edges and spectral gaps through the inequality
\begin{equation} 
\label{eq:3fbounds}
    |E-2V\cos{\left(\phi + m\frac{2\pi}{3}\right)}|< |W|,
\end{equation}
with $m=0,\pm 1$. { As discussed later, the different values of $m$ label the $C_3$ rotation eigenstates of $H_V$. (see also Supp. Mat.\cite{SuppMat}  section \ref{AppendixD}) }

In Fig.~\ref{fig:pd}(a) we compare the energy spectrum as a function of $\phi$ calculated numerically 
using periodic boundary conditions and the Kernel Polynomial Method~\cite{KPM}, with the spectrum outlined by the inequalities~\eqref{eq:3fbounds}. The lines set by \eqref{eq:3fbounds} match exactly with the band edges of the numerical spectrum. The agreement confirms Weaire and Thorpe's original expectation: the local environment of a site is enough to determine the broad spectral features, and where the gap closures appear\cite{Weaire71}. Our goal is to show that these properties also allow us to determine the topological phase diagram. 

To do so we first show that the model can be indeed topologically non-trivial and discuss some of its physical properties. With open boundary conditions we observe that states appear within bulk gaps for certain values of parameters. A typical local density of states of these ingap states is shown in Fig.~\ref{fig:pd}(c). The wave functions of these states are localized at the edge suggestive of a topological edge-mode.

To map the topological phase diagram, and predict physical properties we have calculated the local Chern marker $\mathcal{C}(\mathbf{r})$ at each lattice site $\mathbf{r}$ for different parameter values. The local Chern marker can be regarded as the real space counterpart of the Berry curvature~\cite{Bianco_thesis,LCM}. It is defined at each site as the expectation value {(see Supp. Mat.\cite{SuppMat} section \ref{AppendixA})}
\begin{equation}
\label{eq:LocalChernMarker}
\mathcal{C}(\mathbf{r})= 2\pi\mathrm{Im}\left\langle\mathbf{r} \right|\big[ \hat{Q}\hat{x},\hat{P}\hat{y} \big]\left|\mathbf{r} \right\rangle,
\end{equation}
over localized states $\left|\mathbf{r} \right\rangle$, where $\hat{P}$ and $\hat{Q}$ are projectors onto the occupied and unoccupied eigenstates.

With periodic boundary conditions, and for a two-dimensional insulator, the density of the local Chern marker is equal to the total Chern number, $C = \mathrm{Tr}[\mathcal{C}(\mathbf{r})] / A_{\mathrm{sys}}$, where $A_{\mathrm{sys}}$ is the area of the system~\cite{Bianco_thesis,LCM}. With open boundary conditions $\mathrm{Tr}[\mathcal{C}(\mathbf{r})]=0$, since it is the trace of a commutator in a finite Hilbert space\cite{Bianco_thesis,Tran2017,Pozo2019}. In an atomic insulator $\mathcal{C}(\mathbf{r})$ is zero on all $\mathbf{r}$, resulting in a vanishing trace over all sites. In contrast, when the Chern number is finite the area averaged $\mathcal{C}(\mathbf{r})$ in the bulk takes an integer value that equals $C$, which is compensated exactly by an edge contribution of opposite sign upon averaging over all sites. The local Chern marker thus serves to diagnose topology of 2D insulators without time-reversal symmetry. 

\begin{figure}
    \centering
    \includegraphics[width=\columnwidth]{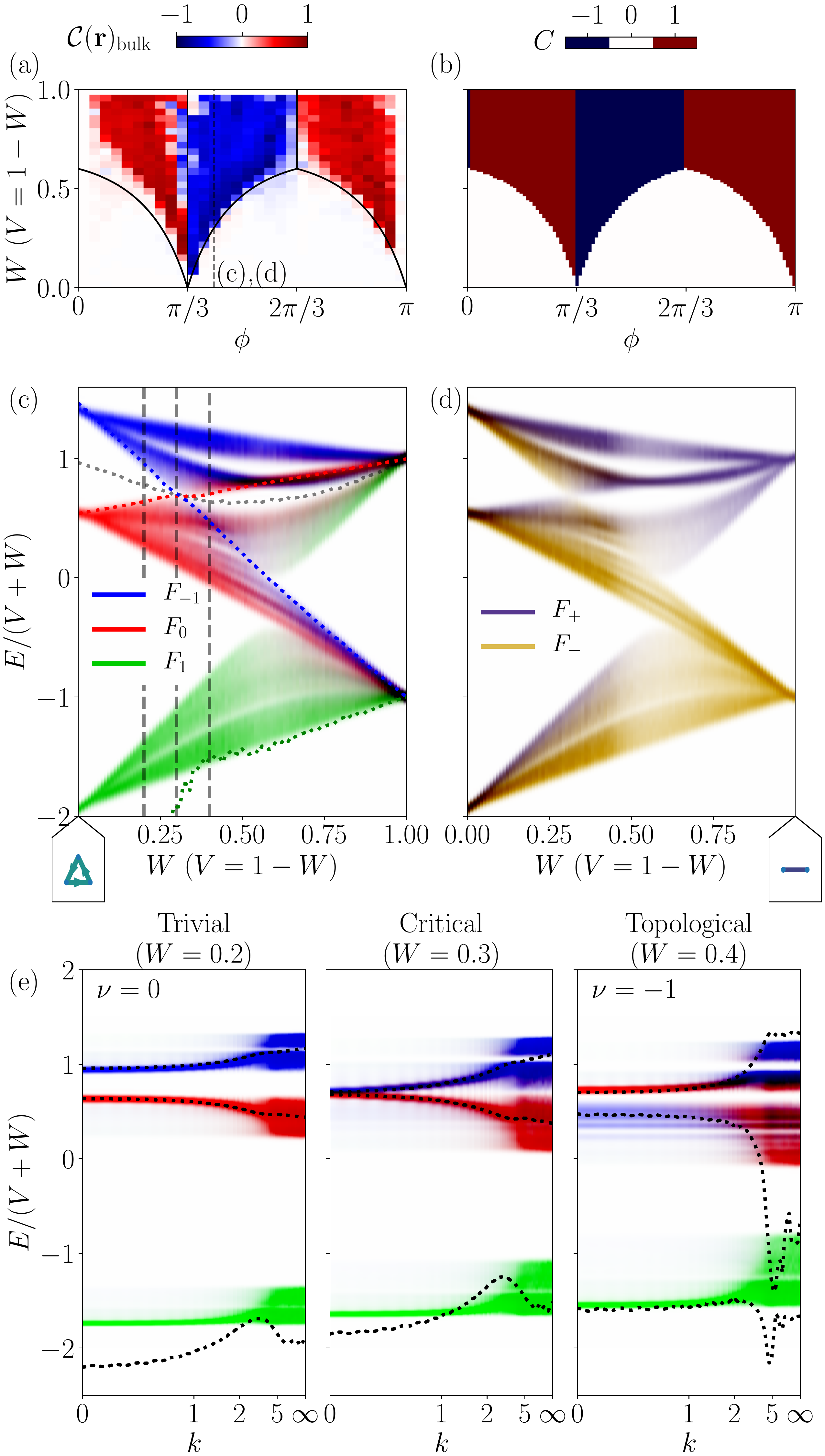}
    \caption{\textbf{Topological phase diagram and symmetry properties { of the $z=3$ Weaire-Thorpe model at $\nu=2/3$ filling}} 
    (a) Topological phase diagram obtained from the local Chern marker density averaged over the area within the dashed white square in Fig.~\ref{fig:pd}(d), $\left\langle c(\mathbf{r})\right\rangle_{\mathrm{bulk}}$. { The vertical dashed line indicates $\phi=1.3$, used in (c) and (d).}
    (b) Topological phase diagram using the symmetry indicator formula Eq.~\eqref{eq:symind}. The solid lines indicate the gap closing transitions obtained using the inequalities Eq.~\eqref{eq:3fbounds}.
    (c) { Spectral densities $F_{m}(E)$. We use an RGB color value to visualize how a given eigenstate tranforms under $C_3$ rotations (see main text).}
    The dotted lines show the effective Hamiltonian spectrum at $\bk=0$ for $l=0,\pm 1$ with the same color coding. The gray dotted line indicates a 2/3 filling. The lower left panel shows the trivial decoupled triangle limit ($W=0$).
    (d) { Spectral densities $F_{\pm}(E)$.  We use a two color coding to visualize how a given eigenstate tranforms under bond inversion (see main text).} The lower right panel shows the trivial dimer limit ($V=0$).
    (e) Momentum resolved spectral weights $F_{m}(E,\mathbf{k})$ showing a band inversion at $|\mathbf{k}|\equiv k=0$. The eigenvalues of $H_\mathrm{eff}(\mathbf{k})$ are shown as dark dotted lines. The continuum Chern number $\nu$ changes from $\nu=0$ to $\nu=-1$ across the transition.}
    \label{fig:symmetry}
\end{figure}

Using the Chern marker we can compute the topological phase diagram, shown in Fig.~\ref{fig:symmetry}(a), for the representative case of $2/3$ filling. It features three phases with Chern numbers $C=0,\pm1$, shown for $\phi \in [0,\pi]$ since $C(\phi) = -C(-\phi)$. 
To visualize quantization in Fig.~\ref{fig:pd}(d) we plot the dimensionless local Chern marker density~\cite{Bianco_thesis} $c(\mathbf{r})$ within the topological state with $C=-1$ at $\phi=1.3$. By definition (see Supp. Matt.\cite{SuppMat} { section \ref{AppendixA}}) $c(\mathbf{r})$ coincides with the density of $\mathcal{C}(\mathbf{r})$ on average, converging to $C$ upon summing over bulk sites. The edge state contribution, with opposite sign, is clearly visible.

The local Chern marker is a tool to elucidate the { topological} phase diagram of this model, but is also connected to physical properties. 
Firstly, the Hall conductivity $\sigma_{xy}$ is determined by the Chern number, $\sigma_{xy}=Ce^2/h$. Secondly, $\mathrm{Tr}[\mathcal{C}(\mathbf{r})]$ determines the absorption rate difference between driving the system with left and right handed circularly polarized electric field of amplitude $E$\cite{Tran2017}. This observation does not rely on translational invariance, and thus our model should show a quantized circular dichroism. By following Ref.~\onlinecite{Tran2017}, we show that the differential frequency-integrated absorption rate is quantized to $\Delta \Gamma/A_{\mathrm{sys}} = E^2\hbar^2 \mathrm{Tr}[\mathcal{C}(\mathbf{r}))]/A_{\mathrm{sys}}=E^2\hbar^2C$ (see Supp. Mat.\cite{SuppMat} { section \ref{AppendixA}}). A finite quantized circular dichroism can be measured even for finite samples, upon integrating  to frequencies up to the band gap\cite{Pozo2019}.\\

%
\noindent\textbf{Symmetry indicators and topological invariants}\\
{ By construction the Weaire-Thorpe models impose that all orbitals are equivalent. We show next how we can use the resulting underlying symmetries, cyclic permutation of the orbitals on a site ($2\pi/z$ rotations) and bond inversion, to determine the topological phase diagram.
For concreteness we consider the $z=3$ case in the following, but our conclusions carry over to the general case with minimal modifications.

The properties of the eigenstates of $H_{\mathrm{WT}}$ under rotations are best understood starting from the trivial limit $W=0$. In this limit, \eqref{eq:Topo_WT} defines a set of decoupled triangles, each governed by $H_V$ (see Fig.~\ref{fig:symmetry}(c) lower left schematic). The system is topologically trivial, since it is possible to form a basis of localized states~\cite{Thouless:1984en,Soluyanov2011}. Its spectrum consists of three bands with $N/3$ states each, where $N$ is the total number of orbitals in the system. Therefore, fillings $N/3$ and $2N/3$ define trivial insulators. Since $H_V$ is invariant under $2\pi/3$ rotations, the local site symmetry group is $C_{3}$. This implies that at $W=0$ all states are exact eigenstates of $C_{3}$, labeled by their rotation eigenvalues $w_m = e^{i 2\pi m/3}$ with $m=0,\pm 1$. 

Crucially the eigenstates of the Hamiltonian $H_{\mathrm{WT}}$ at the band edges determined by Eq.~\eqref{eq:3fbounds} remain eigenvectors of $H_V$ whatever the relative magnitude of $V$, $W$ and $\phi$ (see Supp. Mat.\cite{SuppMat} section \ref{AppendixD} for an analytical derivation). For general states we characterize the transformation properties under threefold rotations by computing $F_m(\ket{\psi}) = \sum_i \vert\braket{i,m|\psi}\vert^2$, the overlap with the eigenvectors $\ket{i,m}$ of $H_V$ localized on site $i$. With respect to $C_3$, $\ket{i,0}$ transforms as an $s$-like orbital with eigenvalue $w_0=1$, while $\ket{i,\pm 1}$ transform as $p_{x}\pm ip_y$-like orbitals with eigenvalues $w_{\pm1}=e^{\pm i 2\pi/3}$. For any state $F_m\geq 0$ and $\sum_m F_m = 1$, so we assign an RGB color code to visualize how it transforms under $C_3$ rotations (see Fig.~\ref{fig:symmetry}(c)). As advertised, states at the band edges have $F_m(\ket{\psi})= 1$ for some $m$ and are exact $C_3$ eigenstates.

Similarly, to understand the properties of the eigenstates of $H_{\mathrm{WT}}$ under inversion we start from the $V=0$ limit. When $V=0$, the system is a set of decoupled dimers (see Fig.~\ref{fig:symmetry}(d), lower right schematic).  The energy spectrum is composed of two bands at energies $\pm W$, with $N/2$ states each, labeled by $\pm 1$ bond inversion eigenvalues. At $1/2$ filling, the system is a trivial insulator. Analogous to our procedure above, we characterize the properties of any eigenstate under inversion away from $V=0$ by introducing  $\ket{j,\pm}$, the eigenvectors of $H_W$ localized on the dimer $j$, and computing $F_\pm(\ket{\psi}) = \sum_j\vert\braket{j,\pm|\psi}\vert^2$. As before, the band edges remain eigenstates of $H_W$ whatever the relative magnitude of $V$, $W$ and $\phi$ (see Supp. Mat.\cite{SuppMat} section \ref{AppendixD}), which can be seen in Fig.~\ref{fig:symmetry}(d). }


Since band edges remain eigenvectors of $H_V$ and $H_W$ separately, and these track band crossings, it is suggestive that using Eq.~\eqref{eq:3fbounds} we can track changes in Chern numbers. This would allow to map a topological phase diagram analytically. { To this end we take inspiration from the idea of symmetry {indicators\cite{Kruthoff17,Bradlyn2017,Po:2017ci,Song:2018cj}} (see Ref.~\onlinecite{Po20} for a review), and extend those developed for Chern insulators~\cite{Fang2012}. Relevant to our analysis, the latter work established in particular that in two-dimensional crystals with $C_{n}$ rotational symmetry, the Chern number can be determined modulo $n$ by multiplying rotation eigenvalues of filled states.  This multiplication amounts to summing the exponents of the filled rotation eigenvalues, $w$ for $C_{3}$, $\sum_{p\in \mathrm{filled}}m_p$. We then observe that} the Chern number at a given point in the phase diagram can be computed as
{
\begin{equation}
\label{eq:symind}
    C (\mathrm{mod}\hspace{0.1cm}3) = \sum_{p\in \mathrm{filled}} m_p - \sum_{p\in \mathrm{filled}}m^{W=0}_p \hspace{0.2cm} .
\end{equation}
}
The second term in this expression acts as a reference for the trivial state{, which is well defined for 1/3 and 2/3 fillings,} while the first tracks band inversions. For $2/3$ filling the resulting phase diagram is shown in Fig.~\ref{fig:symmetry}(b). It reproduces that computed from the local Chern marker (Fig.~\ref{fig:symmetry}(a)), yet its computation is analytical. { A similar invariant can be found for $z=4$ as shown in the Supp. Mat.~\cite{SuppMat}, section \ref{AppendixF}.}


It is appealing to connect the success of the invariant Eq.~\eqref{eq:symind} to known topological invariants. { First, Eq.~\eqref{eq:symind} can be thought as the amorphous analogue of the Chern number equation formula for crystals with $C_3$ symmetry \cite{Fang2012}. Second, in} continuous media,  the Chern number can be computed by subtracting {angular momentum} eigenvalues $l$ of filled states at $|\mathbf{k}|\equiv k=0$ and $k = \infty $~\cite{VanMechelen:2018cy,VanMechelen:2019ha}, by defining $\nu= \sum_{n\in \mathrm{filled}}l_n(k=0) - l_n(k = \infty)$. The $k = \infty$ term captures the short distance properties, { and thus it is suggestive to interpret it as} the second term in Eq.~\eqref{eq:symind}. Similarly, the $k=0$ term captures long distance properties, and { it is tempting to identify it} with the first term Eq.~\eqref{eq:symind}. { Although appealing, this identification is subtle, because even in crystals further neighbour hoppings can break the naive intuition that gap inversions occur at $k=0$. Therefore, } establishing a formal correspondence is an interesting open problem, yet the similarities between $\nu$ and Eq.~\eqref{eq:symind}, and the average rotational symmetry of amorphous lattices suggests that $\nu$ can be used to signal amorphous topological states. 

To investigate this possibility we extend a recent description of topological quasicrystalline phases~\cite{Varjas2019} to our amorphous lattices. By projecting the real space Hamiltonian into a basis of plane waves with a given $\mathbf{k}$ we can define an effective Hamiltonian in momentum space, $H_\mathrm{eff}(\mathbf{k})$\cite{Varjas2019}. Since this procedure does not rely on translational symmetry, we define $H_\mathrm{eff}(\mathbf{k})$ for our amorphous system using a basis of angular momentum states (see Supp. Mat.~\cite{SuppMat}, section \ref{AppendixE}.). 

The symmetry properties of $H_\mathrm{eff}(\bk)$ allow us to compute $\nu$ and compare it to Eq.~\eqref{eq:symind}. 
As $k = 0$ and $k=\infty$ are invariant under continuous rotations, the eigenstates of $H_\mathrm{eff}(k=0)$ and $H_\mathrm{eff}(k=\infty)$ can be labeled by their angular momentum $l$. The colored dotted lines in Fig.~\ref{fig:symmetry}(c) show the eigenstates of $H_\mathrm{eff}(0)$ labeled by $m = l \in [-1, 0, 1]$, which closely follow the spectral densities $F_m (E)$. The $l=0$ and $l=-1$ eigenvalues of $H_\mathrm{eff}(0)$ cross at the first topological phase transition, while the eigenvalues of $H_\mathrm{eff}(\infty)$ maintain the same ordering. This behaviour matches that of the momentum-resolved spectral densities of the permutation eigenstates $F_{m}(E,\mathbf{k})$ (Fig.~\ref{fig:symmetry}(e)) which also present a band inversion at $k = 0$ across the topological transition. For general $\bk$ the eigenstates of $H_\mathrm{eff}(\bk)$ disperse, but remain gapped and continuous, establishing a connection to regularized continuum Hamiltonians\cite{VanMechelen:2018cy,VanMechelen:2019ha}. 
By explicitly computing the invariant $\nu$, that compares the number of filled angular momentum eigenstates at $k = 0$ and $\infty$, we can establish the topological character of this band inversion, which changes $\nu =0$ to $\nu = -1$. 
However, we find that this approach only results in a meaningful $H_\mathrm{eff}(\mathbf{k})$ sufficiently close to the decoupled triangle limit $W/V \lesssim 1$ capturing only part of the phase diagram (see the Supp. Mat.~\onlinecite{SuppMat} section \ref{AppendixE} for a discussion).

\section*{Discussion}

We have proposed a class of realistic models with fixed coordination that allow to analytically track topological phase transitions in amorphous lattices. These models are motivated by the observation that the local environment of a site is similar in the crystalline and amorphous lattice, the latter lacking long-range order. A fixed coordination allows us to show that these models are generically gapped, and the equivalence between orbitals allows us to assign a symmetry label to band edges.  Treating these labels as symmetry indicators we have constructed the topological index~\eqref{eq:symind}, successfully reproducing the topological phase diagram of a threefold Weaire-Thorpe-Chern insulator analytically. We have linked the phase diagram to physical responses, predicting that 2D amorphous models with broken-time reversal symmetry present a quantized circular dichroism, similar to their crystal counterparts~\cite{Tran2017}. The topological index \eqref{eq:symind} can be defined for any $z$, signaling a way to determine the phase diagram of any two-dimensional Weaire-Thorpe model in the Altland-Zirnbauer class $A$ analytically.

Our results are a promising step to incorporate symmetries, such as orbital equivalence or average rotational symmetry, to classify amorphous topological states beyond the tenfold way. These could be combined with the effective Hamiltonian approach~\cite{Varjas2019} and with extra symmetries, such as time-reversal or particle-hole symmetry, to answer the question of whether new topological states, absent in crystals, can exist in amorphous matter. { One way that new phases can appear is by considering local building blocks with symmetries absent in crystals, such as $C_5$ or $C_8$ rotations. It is interesting to speculate if these symmetries could lead to unexpected quantum Hall transitions in amorphous two-dimensional magnetic materials.} Lastly, our models admit easy generalizations to higher dimensions and non-hermitian couplings. 

%
%
Our work establishes that in the absence of translational invariance it is possible to construct topological models that incorporate realistic elements, such as fixed coordination, and for which the topological phase diagram can be computed analytically using symmetry, contrary to naive expectation. They are therefore natural candidates to describe amorphous topological states in the solid state~\cite{Corbae:2019tg}, and they can serve as models for synthetic systems, such as photonic Chern bands, where large optical gaps can be realized using continuous random networks~\cite{Florescu:2009ev,Rechtsman2011}.\\

\noindent\textbf{Acknowledgements}\\
We are grateful to S. Tchoumakov, L. Herviou, and C. Repellin for enlightening suggestions, and P. Corbae, S. Ciocys, E. Dresselhaus, B. Sbierski, A. R. Akhmerov, T. Ojanen, and K. P{\"o}yh{\"o}nen for related collaborations and discussions.  A. G. G. acknowledges financial support by the ANR under the grant ANR-18-CE30-0001-01 and the European Union Horizon 2020 research and innovation programme under grant agreement No 829044. D. V. is supported by
NWO VIDI grant 680-47-53. Our calculations were performed using the Python package $\textsc{kwant}$\cite{Groth_2014} and our plots using $\textsc{matplotlib}$\cite{Hunter:2007}. The code used for the numerical calculations and the data shown in the manuscript is available at Ref.~\onlinecite{zenodo}.\\

\noindent\textbf{Author contributions}\\
Q.~M. performed the analytical calculations and implemented the numerical simulations assisted by D.~V., who developed the connection to the effective Hamiltonian. A.~G.~G. devised the initial concepts and theory, which were further developed by all authors. A.~G.~G. wrote the manuscript, with inputs from Q.~M. and D.~V., and supervised the project.

\newpage

\onecolumngrid

\renewcommand{\thefigure}{S\arabic{figure}}
\setcounter{figure}{0} 

\appendix

\section*{Supplementary Materials}

\section{Local Chern marker and Circular Dichroism}\label{AppendixA}

\subsection{Local Chern marker\label{AppendixA1}}

Two-dimensional insulators in class A of the Altland-Zirnbauer classification are classified by an integer known as the Chern number. In momentum space, the Chern number can be calculated by computing the integral of the Berry curvature of each band over the Brillouin zone, and summing over filled bands~\cite{Haldane04}. To signal a finite bulk Chern number Bianco and Resta \cite{Bianco_thesis,LCM} introduced the local Chern marker $\mathcal{C}(\mathbf{r})$
\begin{equation}
\mathcal{C}(\mathbf{r}) = \sum_l \bra{\mathbf{r}, l}\hat{C}\ket{\mathbf{r}, l},
\end{equation}
where
\begin{equation}
\hat{C} = i\pi \left(\hat{P}\hat{\mathbf{r}}\hat{Q}\times\hat{Q}\hat{\mathbf{r}}\hat{P}-\hat{Q}\hat{\mathbf{r}}\hat{P}\times\hat{P}\hat{\mathbf{r}}\hat{Q}\right),
\end{equation}
defined by the position operator $\hat{\mathbf{r}}$, the projector onto occupied states $\hat{P} = \sum_{n\in \rm occ.} \ket{n}\bra{n}$, and the projector onto unoccupied states $\hat{Q} = \id-\hat{P}$.
The states $\ket{\mathbf{r}, l}$ are localized on a site at position $\mathbf{r}$ in local orbital $l$.
The average density of the local Chern marker over the whole system equals the Chern number, which can be expressed through the trace of the operator $\hat{C}$
\begin{equation}
C = \frac{1}{A_{\rm sys}}\sum_{\mathbf{r}}\mathcal{C}(\mathbf{r}) = \frac{1}{A_{\rm sys}}\mathrm{Tr}(\hat{C}).
\end{equation}
For finite systems this trace is zero, since it is the trace of a commutator in a finite Hilbert space. However, the real space distribution of the local Chern marker signals a Chern insulator state by a quantized value of the Chern marker in the interior of the system, and large and opposite contribution localized at the edges of the sample. Previously this marker has been used to signal finite Chern markers both in crystalline\cite{Bianco_thesis,LCM,Tran2017}  and quasicrystalline systems\cite{Tran:2015cj}, and for our purposes it contains the same information as other real space topological markers, such as the Bott index\cite{Loring:2010jh}.

The local Chern marker $\mathcal{C}(\mathbf{r})$ is a dimensionful quantity that fluctuates as a function of the discrete site positions $\mathbf{r}$, and only its average density corresponds to the dimensionless Chern number $C$ quantized to integers.
In order to visualize the quantized local marker, we follow Ref.~\onlinecite{Bianco_thesis} to define the local Chern marker density $c(\mathbf{r})$ for all positions through convolution with a test function $g(\mathbf{r})$
\begin{equation}
c(\mathbf{r}) = \sum_{\mathbf{r}'} \mathcal{C}(\mathbf{r}') g(\mathbf{r} - \mathbf{r}'),
\end{equation}
where $g(\mathbf{r})$ is chosen to be smooth, circularly symmetric, localized with finite support, and normalized such that $\int  g(\mathbf{r}) d^2\mathbf{r} = 1$.
This ensures that $c(\mathbf{r})$ is dimensionless, its average value coincides with the density of $\mathcal{C}(\mathbf{r})$, and in the limit of macroscopic averaging, when the support of $g(\mathbf{r})$ is much larger than the typical interatomic spacing, it converges to $C$ everywhere.
In particular we choose
\begin{equation}
g(r) = \begin{cases}
      \frac{12}{\pi w^2} \left[1 - \left(\frac{2 r}{w}\right)^2\right]^2 & r\leq w/2, \\
      0 & r > w/2,
   \end{cases}
\end{equation}
where $w$ controls the diameter of the support.
Even for values of $w$ comparable to the interatomic spacing, $c(\mathbf{r})$ becomes a smooth function in the bulk with value fluctuating near $C$, as shown in Fig.\ref{fig:pd}(d) of the main text.

\subsection{Circular Dichroism\label{AppendixA2}}

The local Chern marker is closely related to the total circular dichroism~\cite{Souza2008,Tran2017}, which is the frequency integrated absorption difference between left and right polarizations of an incident electric field, such as that of circularly polarized light. Shining circularly polarized light triggers optical transitions from occupied to unoccupied states depleting the conduction band. We describe the link between the depletion rate to the local Chern marker by following Ref.~\onlinecite{Tran2017}, to explicitly show it does not rely on the periodicity of the lattice. The depletion rate for a given incident electric field of amplitude $E$ is given by the Fermi's Golden Rule
\begin{equation}
    \Gamma_{\mathbf{P}}(\omega) = \frac{2\pi}{\hbar^2}|E|^2\sum_{e\ \in\ \mathrm{unocc}}\sum_{g\ \in\ \mathrm{occ}} |\bra{e}\mathbf{P}\cdot\hat{\mathbf{r}}\ket{g}|^2\delta(\epsilon_e-(\epsilon_g+\hbar \omega)),
\end{equation}
where $\ket{e}$ and $\ket{g}$ are respectively states from the conduction and the valence band, $\mathbf{P}$ is the polarization vector of the electric field and $\omega$ its frequency. For left and right circularly polarized lights, we get
\begin{equation}
    \Gamma_{\pm}(\omega) = \frac{2\pi}{\hbar^2}E^2\sum_{e,g}|\bra{e}\hat{x}\pm i\hat{y}\ket{g}|^2\delta(\epsilon_e-\epsilon_g-\hbar\omega).
\end{equation}
The total integrated differential rate $\Delta\Gamma^{int}$ is given by 
\begin{align}
   \Delta\Gamma^{int} = \frac{1}{2}\int_{0}^{+\infty} (\Gamma_+(\omega)-\Gamma_-(\omega))\diff \omega &= \frac{\pi}{\hbar^2}E^2\sum_{e,g} \label{eq:CD} \bra{g}\hat{x}-i\hat{y}\ket{e}\bra{e}\hat{x}+i\hat{y}\ket{g}-\bra{g}\hat{x}+i\hat{y}\ket{e}\bra{e}\hat{x}-i\hat{y}\ket{g}\\
   &=\frac{2i\pi}{\hbar^2}E^2\mathrm{Tr}\left(\hat{P}\hat{x}\hat{Q}\hat{y}\hat{P}-\hat{P}\hat{y}\hat{Q}\hat{x}\hat{P}\right)\\
   &=\frac{E^2}{\hbar^2}\mathrm{Tr}_\mathbf{r}\mathcal{C}(\mathbf{r}) = \frac{E^2A}{\hbar^2}C.
\end{align}
where $A$ is the area of the system and $C$ is the Chern number, an integer for two-dimensional insulators. In the second to last step we have used that the trace can be expressed in position space, and thus this derivation applies to disordered systems, and in particular to amorphous lattices. 

For a finite system the differential integrated rate vanishes exactly like the trace of the local Chern maker. Experimentally, to measure a quantized circular dichroism, it is necessary to isolate either the edge or the bulk contribution to circular dichroism, since they compensate each other. We review two options to do so, already discussed in the literature~\cite{Tran2017,Pozo2019}. The first, proposed\cite{Tran2017} and implemented in ultra-cold atomic systems~\cite{Asteria:2019if}, is based on a quench protocol. An initial wave-packet is prepared using a confining potential such that it has only a finite overlap with the bulk of the system. After the confining potential is removed the evolution of the total integrated differential rate will be determined by the bulk Chern number $C$. Another alternative is to restrict the frequency integral to be within the bulk-gap. In this frequency window the edge-edge optical transitions dominate over the bulk-edge transitions in the large system size limit~\cite{Pozo2019}, leading to the quantized result $-C$. In practice, since the bulk gap can be unknown, it is sufficient to expand the integration window until a plateau is reached~\cite{Pozo2019}.

\section{Properties of $z-$fold coordinated Weaire-Thorpe models\label{AppendixB}}

\begin{figure}
    \centering
    \includegraphics[width=.8\columnwidth]{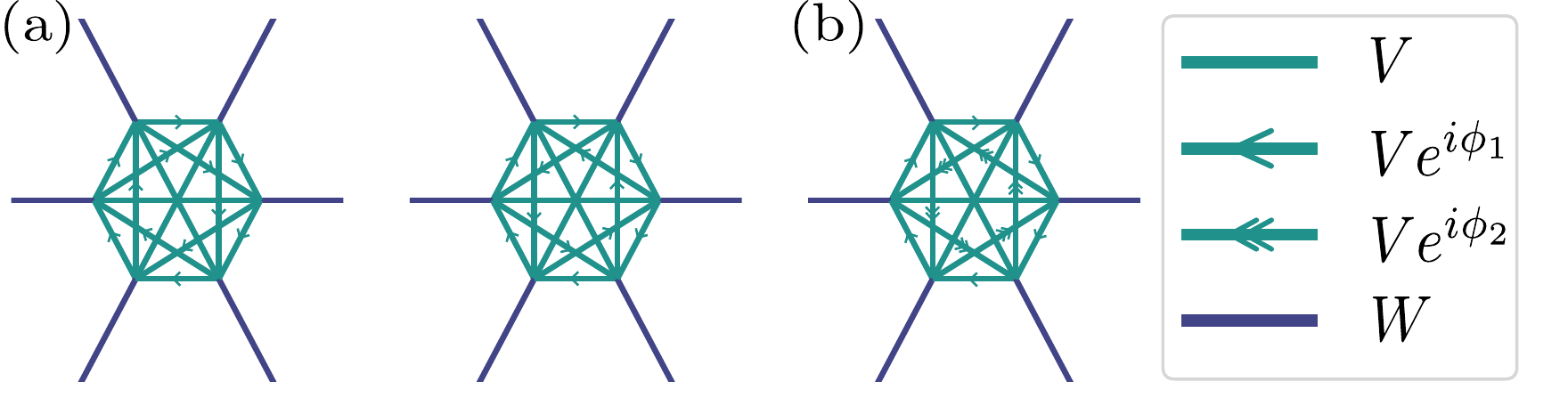}
    \caption{\label{fig:S1poss} Different possible building blocks of a $z=6$ coordinated topological Weaire-Thorpe model that respect the equivalence between orbitals. It is possible to choose the orientation (a) of the phases, and the relative phase (b), while preserving the equivalence between orbitals. The hoppings without arrows have the same magnitude $V$ but must remain real since a $\pi$ rotation would reverse them.}
\end{figure}

The mathematical structure of the topological Weaire-Thorpe models introduced in the main text allows us to derive spectral properties of $z-$fold coordinated lattices. 
As discussed in the main text we allow $H_V$ to have complex hoppings that can induce the appearance of topological phases. 
These models are designed to capture that the local environment of all sites is equal, assuming all orbitals to be equivalent. Accordingly, the hoppings connecting sites are all equal to $W$, while the hoppings in $H_V$ should be invariant with respect to circular permutation of the orbitals. Therefore $H_V$ at a given site is a $z\times z$ matrix of the form
\begin{equation}
H_V=\begin{pmatrix}
0&V_1&V_{2}&\cdots&V_2^*&V_1^*\\
V_1^*&0&V_1&V_{2}&\cdots&V_2^*\\
\vdots&\ddots&\ddots&\ddots&\ddots&\ddots\\
V_1&V_2&\cdots&V_2^*&V_1^*&0
\end{pmatrix}.
\end{equation}
One thus has $\lfloor\frac{z}{2}\rfloor$ coefficients to choose. To retain the equivalence between orbitals, we choose the hoppings to be the same up to a phase. The freedom remaining to choose the phases is restricted by the following considerations
\begin{enumerate}
\item 
The orientation of the phases forming closed hopping loops can be clockwise or counterclockwise so long as they respect the symmetry under permutation of orbitals (see Fig.~\ref{fig:S1poss}(a)). 
\item For even $z$ the phases that can be reversed by a $\pi$ rotation must be real. These are hoppings connecting a given site with the orbital at $z/2$ counting from that site (hoppings without arrows in Fig.~\ref{fig:S1poss}(a) and (b) and Fig.~\ref{fig:models}(a)).
\item For $z >4$ the magnitude of the phase corresponding to the nearest-neighbour hoppings $\phi_1$ (e.g. orbital 1 to orbital 2), the next-nearest-neighbour hoppings $\phi_2$ (e.g. orbital 1 to 3) and so on, are independent in general (see Fig.~\ref{fig:S1poss}(b)). 
\end{enumerate}

Considering the above, the specific phase choice will be determined by the physical system to be represented, yet the quantitative discussion given in the main text still appplies. Specifically we can apply the resolvant method to a general $z-$fold coordinated Weaire-Thorpe Hamiltonian to track the band edges, as shown in Appendix \ref{AppendixC}.

\section{Resolvant method and spectral gaps for $z-$fold coordinated Weaire-Thorpe Hamiltonian\label{AppendixC}}

The Hamiltonian of a $z-$fold coordinated Hamiltonian Weaire-Thorpe model is given by
\begin{equation}
    H = \sum^{z}_{i,j\neq j'} V^{(i)}_{jj'}\ket{i,j}\bra{i,j'} + \sum^{z}_{i\neq i',j} W^{(j)}_{ii'}\ket{i,j}\bra{i',j}.
    \label{eq:Topo_WTSupp}
\end{equation}
To determine the density of states, one can follow Ref.~\onlinecite{Weaire71} to separate the two terms in the hamiltonian as $H = H_V+H_W$, and introduce the resolvent function
\begin{equation}
\varepsilon \mapsto (\varepsilon-H_V-H_W)^{-1}.
\end{equation}
In this decomposition, $H_V$ describes a system of isolated atoms, while $H_W$ is a set of independent dimers. Both $H_V$ and $H_W$ are trivial to solve but their combination makes $H$ non-trivial. The resolvent allows us to determine where are the gaps of the system by noticing that the real poles of the resolvent correspond to the eigenenergies of the Hamiltonian. Therefore, if one can determine the regions of energy where the resolvent is finite, one can show there is no state at these energies. 

The resolvent can be developed into a series
\begin{equation}
(\varepsilon-H_V-H_W)^{-1} = (\varepsilon-H_V)^{-1}\sum_{n=0}^\infty ((\varepsilon-H_V)^{-1}H_W)^n,
\end{equation}
for which a sufficient condition for convergence is $\Vert (\varepsilon-H_V)^{-1}H_W \Vert<1$, where $\Vert \cdot \Vert$ is the operatorial norm given by the maximal absolute eigenvalue of the operator. 

We therefore have $\Vert (\varepsilon-H_V)^{-1}H_W \Vert<\Vert (\varepsilon-H_V)^{-1}\Vert\Vert H_W\Vert$ and
\begin{equation}
\Vert (\varepsilon-H_V)^{-1}\Vert\Vert H_W\Vert<1 \Leftrightarrow \min_{\lambda \in Sp(H_V)}\vert \varepsilon-\lambda\vert>W,
\label{eq:highVcrit}
\end{equation}
where $Sp(H_V)$ runs over the spectrum of $H_V$.
This condition is equivalent to the assertion that states are contained in energy bands whose center are the eigenvalues of $H_V$ and have a bandwidth of $2W$. The eigenvalues of $H_V$ can be simply determined: this operator has $z$ degenerated eigenvalues since it reads $H_V = V\bigotimes I_N$ where N is the total number of sites in the system. 

The criterion \eqref{eq:highVcrit} is useful as long as $W$ is small compared to the distance separating two eigenvalues of $H_V$, but it is not very informative when $W\gg V$. In this latter case, one can develop the Hamiltonian into another series
\begin{equation}
(\varepsilon-H_V-H_W)^{-1} = (\varepsilon-H_W-\overline{V})^{-1}\sum_{n=0}^\infty ((\varepsilon-H_W-\overline{V})^{-1}(H_V-\overline{V}))^n,
\end{equation}
where $\overline{V}$ is a real number chosen such that $\Vert H_V-\overline{V}\Vert$ is minimal.
The convergence criterion now becomes
\begin{equation}\label{eq:lowVcrit}
\min (\varepsilon\pm W-\overline{V})>\max_{\lambda \in Sp(H_V)}(\vert\lambda-\overline{V}\vert).
\end{equation}
The criterion \eqref{eq:lowVcrit} will be informative when W is high compared to the eigenvalues of $H_V$. In this case, we obtain two bands centered in $\pm W+\overline{V}$ whose bandwidth is $2\Vert H_V -\overline{V}\Vert$.

\section{Further details on three-fold coordinated Weaire and Thorpe models\label{AppendixD}}

\subsection{Lattice and Hamiltonian implementation\label{AppendixD1}}

\begin{figure}
    \centering
    \includegraphics[width = \textwidth]{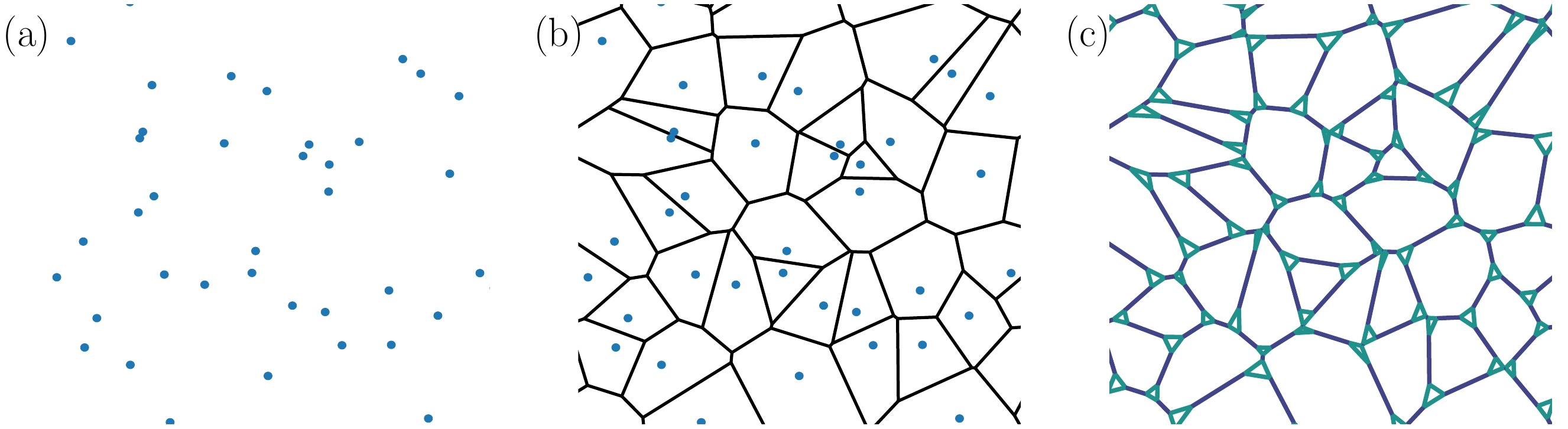}
    \caption{Building steps of a three-fold coordinated tight-binding system. The first step (a) is to plot a random set of points, called seeds. Then, we compute the Voronoi tessellation of the seeds (b): by drawing cells around the seeds such that they gather points that are closer to one seed than any other. { The Voronoi vertices, which are decorated with a triangle, correspond to points at an equal distance to three seeds. Thus, they form the three-fold coordinated lattice discussed in the main text (c).}}
    \label{fig:BuildingSteps}
\end{figure}

To implement the Weaire-Thorpe model in a three-fold coordinated lattice, we first distribute at random a set of points called seeds, and then compute their corresponding Voronoi diagram. The Voronoi diagram is made of Voronoi cells which are defined as regions consisting of all points closer to one seed than to any other. This lattice falls under the continous random network model of amorphous matter, a good model for covalently bonded amorphous solids \cite{Zallen}. The Voronoi vertices will be the sites of our system while the edges of the cells binding them will be the bonds. The building steps of the random trivalent lattice are shown in figure \ref{fig:BuildingSteps}.\\

According to the discussion in Appendix \ref{AppendixB}, for each triangle we choose a clockwise orientation of the complex hopping within each triangular plaquettes. The resulting $H_V$ in \eqref{eq:Topo_WT} is
\begin{equation}
    H_V = V\begin{pmatrix}
    0 & e^{i\phi} & e^{-i\phi}\\
    e^{-i\phi} & 0 & e^{i\phi}\\
    e^{i\phi} & e^{-i\phi} & 0\\
    \end{pmatrix},
\end{equation}
with eigenvalues $\lambda_0 = 2V \cos{\phi}$, $\lambda_{\pm1} = 2V\cos{\left(\phi \pm \frac{2\pi}{3}\right)}$. The Hamiltonian \eqref{eq:Topo_WT} is composed of the above onsite interaction between orbitals of the the same site $H_V$ and an intersite  term $H_W$ of constant hoppings $W$. Due to the trivalent structure of the sites composing our amorphous system the phase diagrams we obtain are $2\pi/3$-periodic.

\subsection{Symmetries of the eigenstates\label{AppendixD2}}

Both $H_V$ and $H_W$ have useful symmetries which we exploit in the main text. Indeed, $H_W$ commutes with the bond inversion $I$ and {$H_V$ for coordination $z=3$ commutes with the generator of $2\pi/3$ rotations, $C_3$}. For the specific case $z=3$, $I$ and $C_3$ read
\begin{equation}
    I = \bigoplus_{\mathrm{bonds}\ j}\begin{pmatrix}
    0&1\\
    1&0
    \end{pmatrix}, \hspace{2mm}
   { C_3 = \bigoplus_{\mathrm{sites}\ i} \begin{pmatrix}
    0&1&0\\0&0&1\\1&0&0
    \end{pmatrix},}
\end{equation}
Therefore, the eigenstates of $H_V$ (respectively~$H_W$) also are eigenstates of {$C_3$} (respectively~$I$). One can then introduce $(\ket{i,m})_{m\in \set{0,\pm 1}}$, the eigenstates of $H_V$ and {$C_3$} localized on triangle $i$, such that
\begin{eqnarray}
H_V\ket{i,m} &=& 2V\cos \left(\phi+m\frac{2\pi}{3}\right)\ket{i,m} \equiv \lambda_m \ket{i,m},\\
\label{eq:xirot}
{ C_3\ket{i,m}} &=& e^{i\frac{2m\pi}{3}}\ket{i,m} {\equiv w_m \ket{i,m}} .
\end{eqnarray}
Similarly, for $H_W$ we introduce the eigenstates $\ket{j,\pm}$ localized on bond $j$ as
\begin{eqnarray}
H_W\ket{j,\pm} &=& \pm W\ket{j,\pm},\\
I\ket{j,\pm} &=& \pm \ket{j,\pm}.
\end{eqnarray}
The eigenstates of $H$ are not, in general, eigenstates of $H_V$ nor of $H_W$. However, it is useful to define their average local similarity~\cite{Weaire71} that we define as
\begin{equation}
F_m(\ket{\psi}) = \sum_i \lvert\braket{i,m|\psi}\rvert^2 = \bra{\psi} \hat{F}_m \ket{\psi}
\end{equation}
and
\begin{equation}
F_\pm(\ket{\psi}) = \sum_j\lvert\braket{j,\pm|\psi}\rvert^2 = \bra{\psi} \hat{F}_{\pm} \ket{\psi}
\end{equation}
where we introduced the operators $\hat{F}_m = \sum_i \ket{i,m}\bra{i,m}$ and $\hat{F}_{\pm} = \sum_j \ket{j, \pm}\bra{j, \pm}$. { As mentioned in the main text with respect to $C_3$, $F_{0}$ transforms as an $s$-like orbital, while $F_{\pm 1}$ transforms as $p_{x}\pm ip_y$-like orbital. Similarly, $F_{\pm}$ distinguish $\pm1$ inversion eigenvalues of a dimer with respect to its bond center, also referred to as 
antisymmetric
($-$) and 
symmetric
states ($+$).}

The values of these similarities are constrained by the normalization of the wave function. Indeed, since $\ket{i,m}$ and $\ket{j,\pm}$ are orthonormal bases of the Hilbert space (the latter is a basis only if the system has no edges, i.e. it has periodic boundary conditions or in the thermodynamic limit), one has
\begin{eqnarray}
\label{eq:ortho1}
F_0 + F_1+F_{-1} = \braket{\psi|\psi} = 1,\\
\label{eq:ortho2}
F_+ + F_- = \braket{\psi|\psi} = 1.
\end{eqnarray}
In order to study properties of systems in the thermodynamic limit, it is useful to introduce the spectral density of these operators as
\begin{equation}
F_m(E) = \frac{1}{N}\mathrm{Tr}\left[\delta (H - E) \hat{F}_m \right]
\end{equation}
and similarly for $F_{\pm}(E)$.
In this case the sum rule is modified such that the total is the full density of states:
\begin{equation}
\sum_m F_m(E) = \rho(E) \equiv \frac{1}{N}\mathrm{Tr}\left[ \delta(H - E) \right].
\end{equation}
These quantities are efficiently calculated using the Kernel Polynomial Method~\cite{weisse2006kernel,Varjas2019b} and are shown on Fig.~\ref{fig:symmetry} (c) and (d).

Another relation can be obtained by projecting the Schr\"{o}dinger equation onto the bra $\bra{\psi}$
\begin{equation}
\label{eq:Scrho}
    \braket{\psi|H_V-E|\psi} = -\braket{\psi|H_W|\psi}.
\end{equation}
The left hand side can be expanded in the eigenbasis of $H_V$, resulting in $\sum_m (\lambda_m-E) F_m$. The right hand side can be expanded in the eigenbasis of $H_W$, resulting in $-W(F_+-F_-)$. These manipulations lead to the relationship
\begin{equation}
\label{eq:cond3}
    \sum_m (\lambda_m-E)F_m = -W(F_+-F_-).
\end{equation}
A final relation between the similarities is given by the square of the Schr\"{o}dinger equation
\begin{equation}
\braket{\psi|(H_V-E)^2|\psi} = \braket{\psi|H_W^2|\psi}.
\end{equation}
As before, expressing each side in its eigenbasis gives 
\begin{equation}
    \label{eq:cond4}
    \sum_{m=-1}^1 \left(\frac{\lambda_m-E}{W}\right)^2F_m = 1.
\end{equation}
Collecting \eqref{eq:ortho1}, \eqref{eq:ortho2},  \eqref{eq:cond3} and \eqref{eq:cond4} we obtain a set of four equations
\begin{eqnarray}
F_0+F_1+F_{-1} = 1, \label{eq:Fs=1}\\ 
F_++F_- = 1, \label{eq:Fpm=1}\\
a_0F_0 + a_1F_1+a_{-1}F_{-1} = F_--F_+, \label{eq:Fpm=Fs}\\
a_0^2F_0+a_1^2F_1+a_{-1}^2F_{-1} = 1, \label{eq:aFs=1}
\end{eqnarray}
where $a_m = \frac{\lambda_m-E}{W}$.

Even though one last equation is required to solve this system of equations exactly, this system constrains the local averaged similarities. Indeed, numerical computations show that their exact value depends on the specific structure of the system. However, these four equations already determine partly the symmetries of the eigenstates as a function of energy. 

Let us focus for example on $F_0$, $F_1$, $F_{-1}$, set by equations \eqref{eq:Fs=1} and \eqref{eq:aFs=1} and represent the system in a 3-D space, each dimension representing one of the $F_m$. Since all $F_m$ remain in the interval $[0,1]$, \eqref{eq:Fs=1} and \eqref{eq:aFs=1} each constrain the solutions to be on a triangle whose vertices lay at $(1,0,0)$, $(0,1,0)$, $(0,0,1)$ and $(1/a_0^2,0, 0)$, $(0,1/a_1^2,0)$, $(0,0,1/a_{-1}^2)$, respectively. Therefore the solutions to equations \eqref{eq:Fs=1} and \eqref{eq:aFs=1} lay on the segment at the intersection of these two triangles. 
If the system is not degenerate, at the band edges set by \eqref{eq:lowVcrit} or \eqref{eq:highVcrit}, one and only one of the $a_m$ is $\pm 1$. Therefore, the segment representing the solutions of \eqref{eq:Fs=1} and \eqref{eq:aFs=1} shrinks into a single point that is $F_m = 1$ and $F_{m'\neq m} = 0$. Coming back to equations \eqref{eq:Fpm=1} and \eqref{eq:Fpm=Fs} then gives $F_+ = 1$ and $F_- = 0$, or vice versa, depending on the sign of $a_m$.

As a consequence, on the edges of the bands determined in \eqref{eq:lowVcrit} or \eqref{eq:highVcrit}, the eigenstates of $H$ remain eigenstates of both $H_V$ and $H_W$ whatever the values of $V$, $W$, or $\phi$ with the eigenvalues summed up in table \ref{tbl:xpctF}. For intermediate energies, the values of $F_m$ {interpolate between one vertex of the triangle defined by \eqref{eq:Fs=1} to another as a function of the energy.} These results can be confirmed numerically as shown in Fig.~\ref{fig:symmetry}(c) and (d).

\begin{table}
\centering
\begin{tabular}{C|C|C|C|C|C|C}
     E& \lambda_0+W & \lambda_0-W & \lambda_{+1}+W & \lambda_{+1}-W & \lambda_{-1}+W & \lambda_{-1}-W  \\
     \hline
     F_m& F_0 = 1 & F_0 = 1 & F_{+1} = 1 & F_{+1} = 1 & F_{-1}=1 & F_{-1} = 1\\
     F_\pm& F_+ = 1 & F_- = 1 & F_+ = 1 & F_- = 1 & F_+ = 1 & F_- = 1
\end{tabular}
\caption{For each of the energies given in the first row, one of the $a_m$ is $\pm 1$ and thus the system \eqref{eq:Fs=1}-\eqref{eq:aFs=1} has a unique solution. The two last rows then indicate which one of $F_m$, $F_\pm$ are $1$, the rest being $0$.}
\label{tbl:xpctF}
\end{table}

\subsection{Sixfold rotation eigenvalues}

\begin{figure}
    \centering
    \includegraphics[width =\columnwidth]{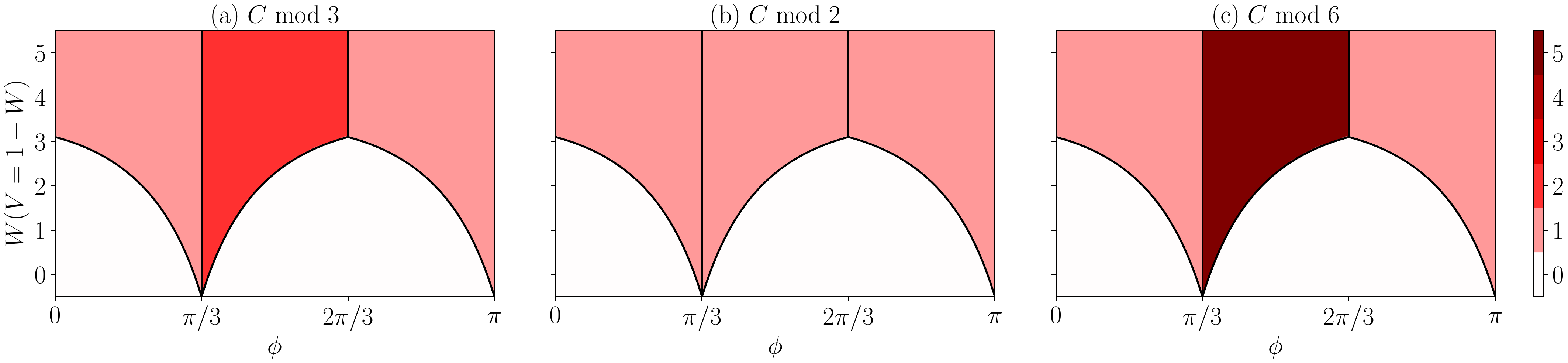}
    \caption{Chern number $C$ as a function of $\phi$ calculated (a) modulo 3 as in the main text, (b) modulo 2, as determined by 
    inversion eigenvalues, 
    and (c) modulo 6, using \eqref{eq:symind6}. All are consistent with the local Chern marker calculation shown in Fig.~\ref{fig:symmetry}(a).}
    \label{fig:Cmod6}
\end{figure}

The formula to calculate the Chern number from the angular momentum eigenvalues of occupied states proposed in the main text, and inspired by Refs.~\onlinecite{VanMechelen:2018cy,VanMechelen:2019ha}, assumes knowledge of the transformation properties under continuous rotations.
Under a rotation by arbitrary angle $\theta$ a state with angular momentum $l$ acquires a phase $\exp{(i \theta l)}$.
In our system, however, through the expectation values $F_0$, $F_{\pm1}$ we only have access to the transformation properties under a permutation of the vertices of a triangle, { corresponding to a threefold rotation with $\theta = 2 \pi /3$ ($C_3$) and rotation eigenvalues $w_m = \exp(\frac{2\pi i}{3} m)$}. This only gives information about the angular momentum modulo 3
\begin{equation}
l = m \mod 3,
\end{equation}
hence the modulo in \eqref{eq:symind}.
On the other hand, {$F_{\pm}$ correspond to the $+1$ and $-1$ inversion eigenvalues or} twofold rotations around the bond centers, and provide information about the angular momentum modulo 2
\begin{equation}
l = n \mod 2,
\end{equation}
where $n = 0, 1$ correspond to $+1$ and $-1$ eigenvalues respectively.
Combining these two equations we can reconstruct $l$ more accurately
\begin{equation}
l = -2 m - 3 n \mod 6.
\end{equation}
A consistency check is to calculate the corresponding Chern number formula
\begin{equation}
\label{eq:symind6}
    C = \sum_{p\in \mathrm{filled}} l_p - \sum_{p\in \mathrm{filled}}l^{W=0}_p \hspace{0.2cm} \mathrm{mod}\hspace{0.1cm}6,
\end{equation}
which is consistent with our numerical phase diagram based on the local Chern marker and Eq.~\eqref{eq:symind} as shown in Fig.~\ref{fig:Cmod6}.

\section{Effective Hamiltonian invariant\label{AppendixE}}
\subsection{Properties of the effective Hamiltonian\label{AppendixE1}}
To obtain an effective continuum $k$-space Hamiltonian we use the construction of Ref.~\onlinecite{Varjas2019}.
We define the effective Hamiltonian $H_{\rm eff}(\bk) = G_{\rm eff}(\bk)^{-1} + E_F$ through the projection of the single-particle Green's function onto plane-wave states
\begin{equation}
\label{eq:Geff}
G_{\rm eff} (\bk)_{l,l'}  = \bra{\bk, l} G \ket{\bk, l'},
\end{equation}
where $G = \lim_{\eta\to 0} \left( H - E_F + i \eta\right)^{-1}$ is the Green's function of the full Hamiltonian with $E_F$ chosen to be in a gap.
The states $\ket{\bk, l}$ are normalized plane-wave states with angular momentum $l$ on the triangles, given in the real space basis by
\begin{equation}
\label{eq:plane_wave}
\braket{\br| \bk, l} = \frac{1}{\sqrt{N}}\exp(i \bk \br_c) \exp(i \phi_{\br} l),
\end{equation}
where $\br$ is the position of a site on the decorated lattice, $\br_c$ is the Voronoi center position in the triangle the site belongs to, $\phi_{\br}$ is the angle of $\br - \br_c$ relative to the $x$ axis (in a clockwise sense), and $N$ is the number of sites in the sample.
Note that this set of states is different from a linear combination of the permutation eigenstates $\ket{i, m}$, as the phases depend on the shape of the triangle in question.
States with different $l \in \mathbb{Z}$ are orthogonal in the thermodynamic limit of $N \to \infty$ because of the averaging over triangles with uniformly distributed orientation.
The basis is, however, overcomplete with respect to $\bk$, because the overlap between different $\bk$ states with the same $l$ only decays as $1/\sqrt{N}$ when approaching the thermodynamic limit.

A central property of $H_{\rm eff}$ is that its gap closes only when the gap of the full Hamiltonian closes.
This follows from the fact that $H_{\rm eff} - E_F$ can only have a zero if $G_{\rm eff}$ has a pole, which is only possible if $G$ has a pole, when $H - E_F$ has a zero.
Hence, a topological invariant defined in terms of $H_{\rm eff}$ that can only change when its gap closes is also a good topological invariant for the original system.
In the large $|\bk|\equiv k$ limit the expectation value in \eqref{eq:Geff} reduces to purely on-triangle terms, as the relative phases between different triangles average to $0$ in the thermodynamic limit, resulting in $H_{\rm eff}(k=\infty)$ being identical to $H_{\rm eff}^{W=0}(k=0)$ in the system with $W$ set to zero.
The limit $\lim_{k \to \infty} H_{\rm eff}(k\hat{n}) \equiv H_{\rm eff}(|\mathbf{k }| = \infty)$ is independent of the direction of the unit vector $\hat{n}$, which allows compactification of $\bk$-space to a sphere.
In practice we construct the $k=\infty$ state using independent random phases on each triangle.

Assuming that $H_{\rm eff}(\bk)$ is finite, gapped, and continuous for all $\bk$, this construction provides a mapping between infinite amorphous Hamiltonians and continuum Hamiltonians.
In the thermodynamic limit the effective Hamiltonian (also the effective Green's function) is invariant under continuous rotations
\begin{equation}
H_{\rm eff}(\bk) = U_{\theta} H_{\rm eff}\left(R_{\theta}^{-1} \bk \right) U_{\theta}^{-1},
\end{equation}
where $R_{\theta}$ is a two-dimensional rotation matrix with a clockwise angle $\theta$, and $\left(U_{\theta}\right)_{ll'} = \delta_{ll'} \exp(i \theta l)$ is the angular momentum representation in this basis.
The momenta $\bk=0$ and $\bk=\infty$ are invariant under rotations, hence $H_{\rm eff}(\bk)$ is diagonal for these momenta in the thermodynamic limit.
This allows to assign definite angular momentum eigenvalues to all eigenstates at these momenta, and use the continuum formula for the Chern number~\cite{VanMechelen:2018cy,VanMechelen:2019ha}
\begin{equation}
\label{eq:cont_chern}
    \nu = \sum_{n\in \mathrm{filled}} l_n(k=0) - \sum_{n\in \mathrm{filled}}l_n(k=\infty),
\end{equation}
where $l_n(\bk)$ are the angular momentum eigenvalues of the filled eigenstates at $k = 0, \infty$.
Strictly speaking this formula is valid when all $l \in \mathbb{Z}$ angular momentum states are taken into account.
We argue that large $l$ eigenvalues do not invert between $k=0$ and $\infty$ and can be safely ignored. This is because the rapid bond-direction dependence of the phase of inter-triangle $W$ hoppings leads to them averaging to zero even at $k = 0$, making the effective Hamiltonian matrix elements identical at $k=0$ and $k=\infty$.

A subtle issue with this construction is that some eigenvalues of $G_{\rm eff}$ might cross zero, even whithout any discontinuous change in the full $G$.
When this occurs, some eigenvalues of $H_{\rm eff}$ diverge, and might move to the other side of $E_F$ without ever crossing $E_F$, see Fig.~\ref{fig:Heff}(a).
In this case, the effective Hamiltonian fails to provide a meaningful continuum model.
We argue that this construction is still applicable in a finite vicinity of the triangle limit ($W/V\ll 1$), where it provides a continuous mapping between infinite amorphous Hamiltonians and continuum $\bk$-space Hamiltonians.
The extent of this region depends on the choice of the $l$ states included in the effective Hamiltonian.
Limiting the set of $l$'s considered might result in some gap closings in the full $H$ being absent in $H_{\rm eff}$, as well as the Chern number formula giving incorrect results if $l$ states that are inverted between $k=0$ and $k=\infty$ are excluded.
However, the gap closings that do appear in $H_{\rm eff}$ with a given set of $l$'s unambiguously signal gap closings, hence serve as indicators of potential topological phase boundaries in the amorphous model.

\subsection{Effective Hamiltonian for the three-fold coordinated topological Weaire-Thorpe-Chern model\label{AppendixE2}}

As shown in Fig.~\ref{fig:symmetry}~(c), the phase transition between the trivial $\nu = 0$ and topological $\nu = -1$ phase is accompanied by an inversion of the $l = 0$ and $-1$ bands of $H_{\rm eff}$ at $|\bk| = 0$.
To further clarify the nature of this band inversion, we also define a related quantity, the $\bk$ and $l$-resolved spectral function
\begin{equation}
A(\bk, l, E) = \bra{\bk, l} \delta(H - E) \ket{\bk, l}
\end{equation}
where $\ket{\bk, l}$ are the states defined in \eqref{eq:plane_wave}.
As shown in Fig.~\ref{fig:symmetry}~(e) the weights of the spectral function closely follow the spectrum of the effective Hamiltonian eigenvalues, and show a band inversion across the phase transition resembling that of crystalline systems.

\begin{figure}
    \centering
    \includegraphics[height=2.2in]{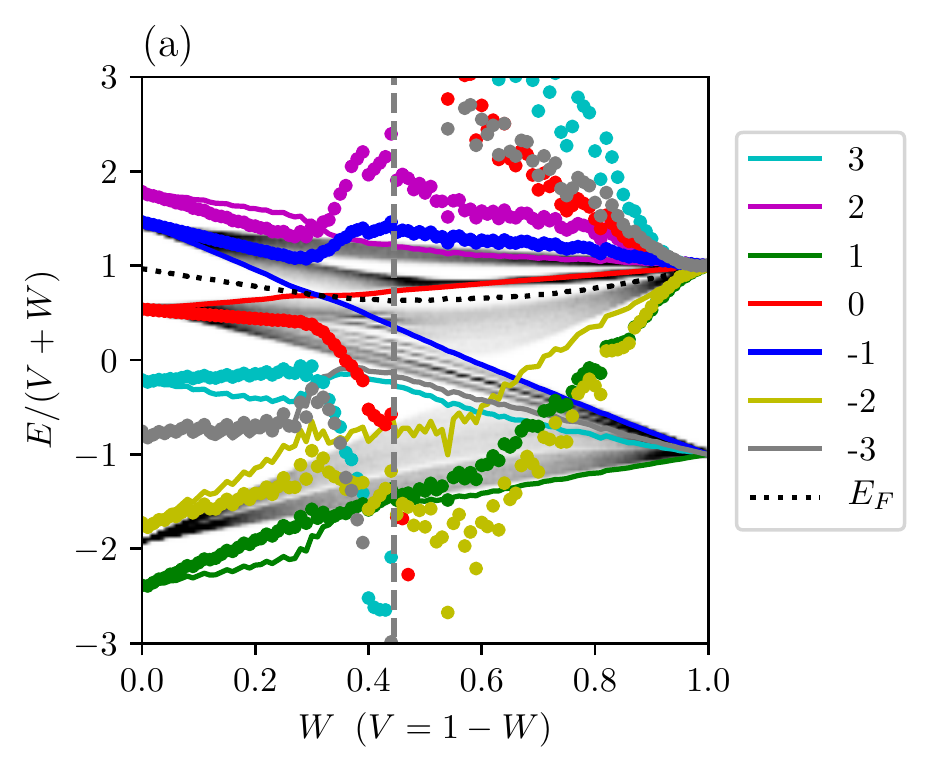}
    \includegraphics[height=2.2in]{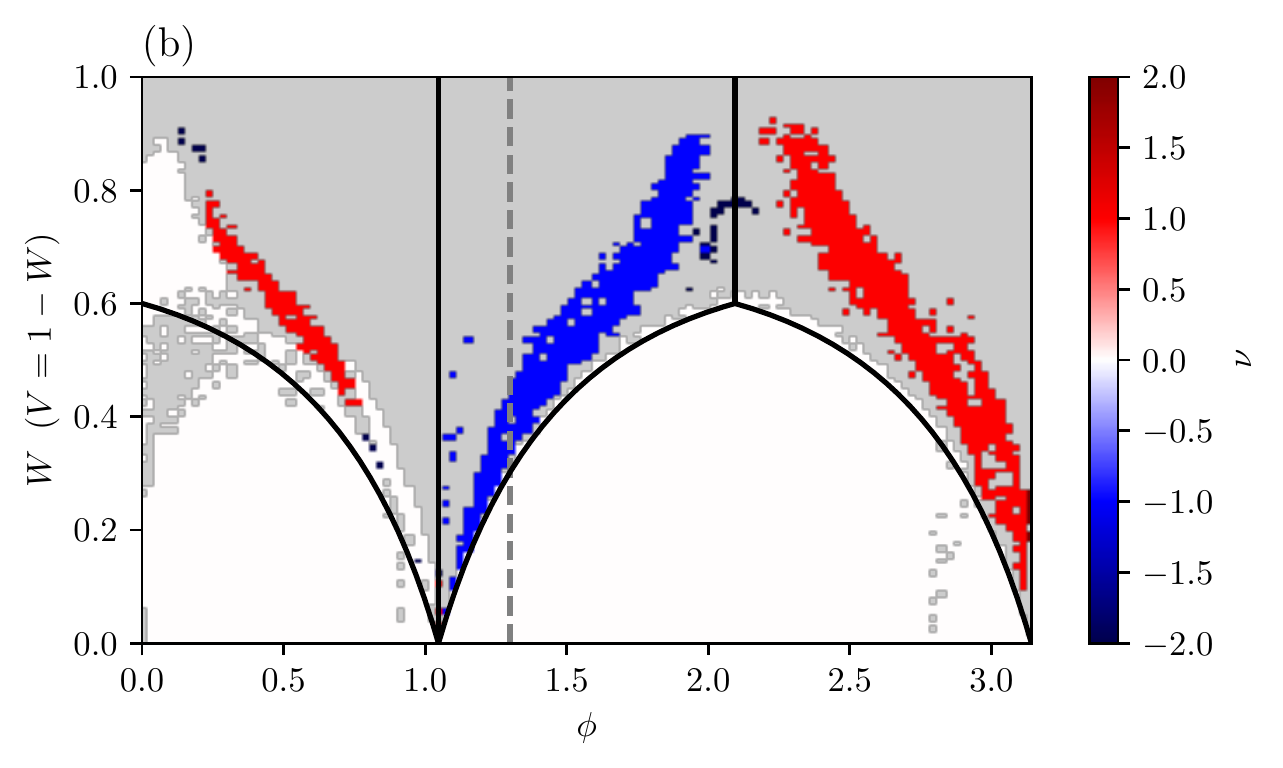}
    \caption{(a) Eigenvalues of the effective Hamiltonian $H_{\rm eff}(k=0)$ (lines) and $H_{\rm eff}(k=\infty)$ (dots) as function of $W$ for $\phi=1.3$ using angular momentum states $l = [-2, \ldots, 3]$.
    The Fermi level is indicated in dotted line and the total density of states is shown in greyscale.
    Some eigenvalues of $H_{\rm eff}(k=\infty)$ diverge (vertical dashed line), and switch sign, the effective Hamiltonian is discontinuous for larger $W$. (b) Chern number $\nu$ as a function of $\phi$ and $W$ calculated using the effective Hamiltonian with $l = [-2, \ldots, 3]$.  In the grey regions the Chern number cannot be determined by this method.     The vertical dashed line shows $\phi = 1.3$.}
    \label{fig:Heff}
\end{figure}

To investigate the full phase diagram, we include $l = [-2, \ldots, 3]$ states and only calculate $H_{\rm eff}$ at $k=0$ and $k=\infty$, sufficient to evaluate \eqref{eq:cont_chern}.
We choose the Fermi level $E_F$ to be in the middle of the gap at $2/3$ filling.
If the number of occupied bands in $H_{\rm eff}$ is different at $k=0$ and $k=\infty$ we conclude that the procedure did not succeed and leave the Chern number undetermined.
The resulting partial phase diagram is shown in Fig.~\ref{fig:Heff}(b), showing transitions between the $W \ll V$ trivial and the neighboring $\nu=\pm 1$ regions.

As argued in this Appendix, the effective Hamiltonian formalism introduced in Ref.~\onlinecite{Varjas2019} provides useful insights to the electronic structure and phase diagram of the topological Weaire-Thorpe class models.
The precise range of applicability of this method in general, however, remains unclear and future work is required to mitigate the issues detailed above.

{
\section{Fourfold coordinated Weaire-Thorpe model\label{AppendixF}}
\subsection{Construction of fourfold coordinated random planar graphs}
We start by generating a set of random lines in the plane using the construction of Ref.~\onlinecite{Miles1964}.
Each line is given by its offset from the origin $p$ and its angle $\theta$, where $p$ is chosen according to a Poisson point process on the positive real line with rate $\tau = 2 \sqrt{\pi \rho}$ and $\theta \in [0, 2\pi)$ is uniformly distributed.
Treating the intersection points as vertices and the line segments between them as links, this results in a homogeneous and isotropic ensemble of strictly fourfold coordinated planar graphs on the plane, with density $\rho$ of vertices~\cite{Miles1964}.

To generate finite samples representative of this ensemble in the circle of radius $R$ around the origin, we choose the number of lines according to the Poisson distribution with mean $2 R \sqrt{\pi \rho}$, and pick the offsets as independent uniformly distributed random variables in $[0, R]$.
We truncate the resulting graph to the interior of the radius $R$ circle, and perform relaxation of the structure to decrease the bond-length fluctuations.

\subsection{Numerical results}
We implement the $z=4$ Weaire-Thorpe model sketched in Fig.~\ref{fig:models}~(a) on random planar graphs, and perform a similar analysis to the $z=3$ case. We show the density of states for $V=W$ as a function of the hopping phase $\phi$ for a finite sample in Fig.~\ref{fig:fourfold_phi}(a). In Fig.~\ref{fig:fourfold_phi}(b) and (c) we show the spectral densities of the projectors onto the onsite fourfold rotation eigensubspaces $F_l(E)$ for $l \in \{0, 1, 2, 3\}$ and onto the subspaces even and odd under inversion, $F_{\pm}(E)$, as function of $W=1-V$ at fixed $\phi = 2\pi /3$.

Similar to the $z=3$ case, inversions between band edges with different symmetry eigenvalues occur at the topological phase transitions.
We find several topological gaps when tuning the hopping phase $\phi$ at $V=W$, with in-gap states localized at the edges of the system (see Fig.~\ref{fig:fourfold_chern}(a)). We calculate the topological phase diagram at half filling (Fig.~\ref{fig:fourfold_chern}(c)) using the averaged local Chern marker density, and show a representative real space distribution when the parameters are chosen in a topological phase in Fig.~\ref{fig:fourfold_chern}(b).

\subsection{Symmetry indicator Chern number formula}

In analogy with the $z=3$ case we can write a symmetry indicator that delivers the Chern number modulo 4.
At half-filling W=0 describes a trivial insulator, that we use as a reference. The band edges can be labeled by eigenvalues of $C_4$, which are $\xi_l = e^{2\pi il/4} = \pm1, \pm i$. In terms of $l$ of the filled states the Chern number is given by

\begin{equation}
\label{eq:symind4}
    C (\mathrm{mod}\hspace{0.1cm}4) = \sum_{p\in \mathrm{filled}} l_p - \sum_{p\in \mathrm{filled}}l^{W=0}_p \hspace{0.2cm} .
\end{equation}

By combining this equation with the expressions for the band edges obtained by the resolvent method described in the main text and section \ref{AppendixC} we obtain the analytical topological phase diagram shown in Fig.~\ref{fig:fourfold_chern}(d). As with $z=3$, it captures the features of the phase diagram obtained by computing the averaged local Chern marker, shown in Fig.~\ref{fig:fourfold_chern}(c).

\begin{figure}
    \centering
    \includegraphics[width = \textwidth]{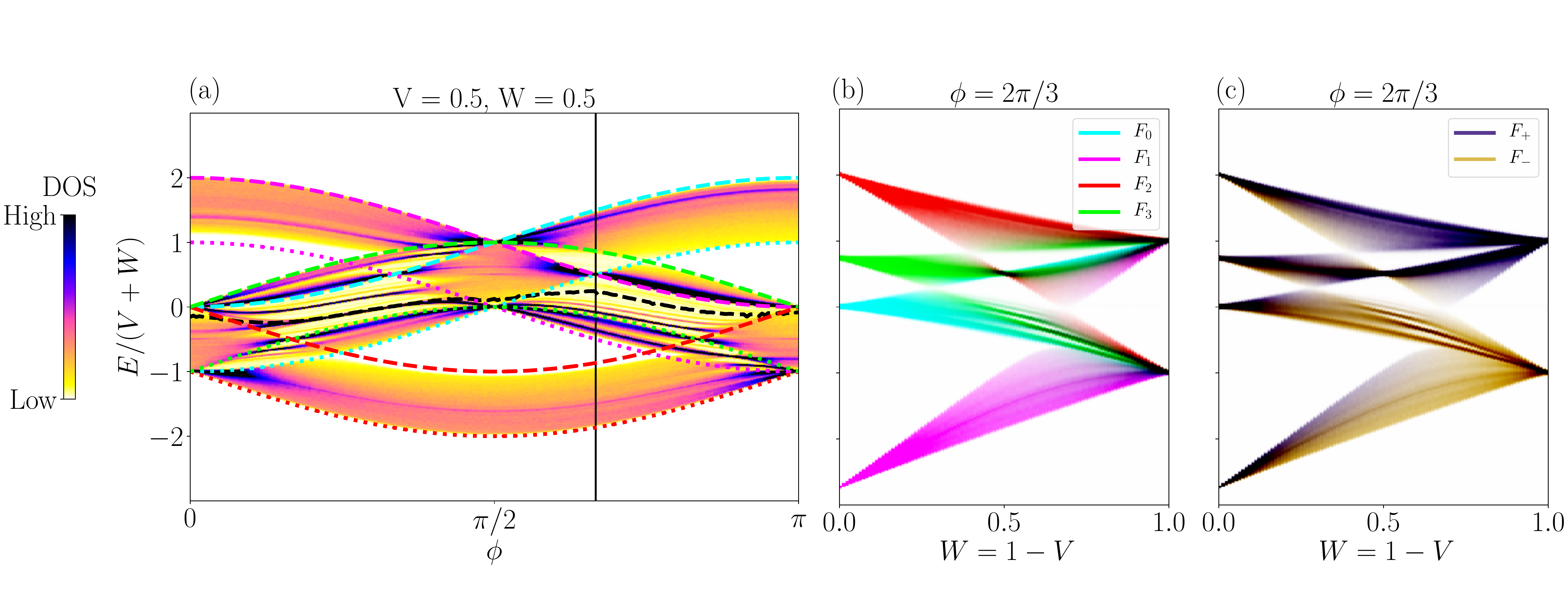}
    \caption{(a) Density of states for a finite sample of the fourfold coordinated WT model with radius $R=30$ and density $\rho=1$.
    Several topological gaps are visible with small finite density of states that is due to topological edge states (see Fig.\ref{fig:fourfold_chern})(a)).
    The dashed black line shows the Fermi level at half filling.
    The vertical line shows $\phi = 2\pi/3$ which we fix for the rest of the figures.
    The dotted and dashed coloured lines correspond to the band edges determined analytically with the resolvent method.
    (b) Spectral density of the projectors onto the onsite eigenstates of the system as a function of the parameters $V$ and $W$ of the Hamiltonian. 
    (c) Spectral density of the projectors onto the bond eigenstates of the lattice. $F_\pm$ project onto states with $\pm1$ inversion eigenvalues.}
    \label{fig:fourfold_phi}
\end{figure}

\begin{figure}
    \centering
    \includegraphics[width = .8\textwidth]{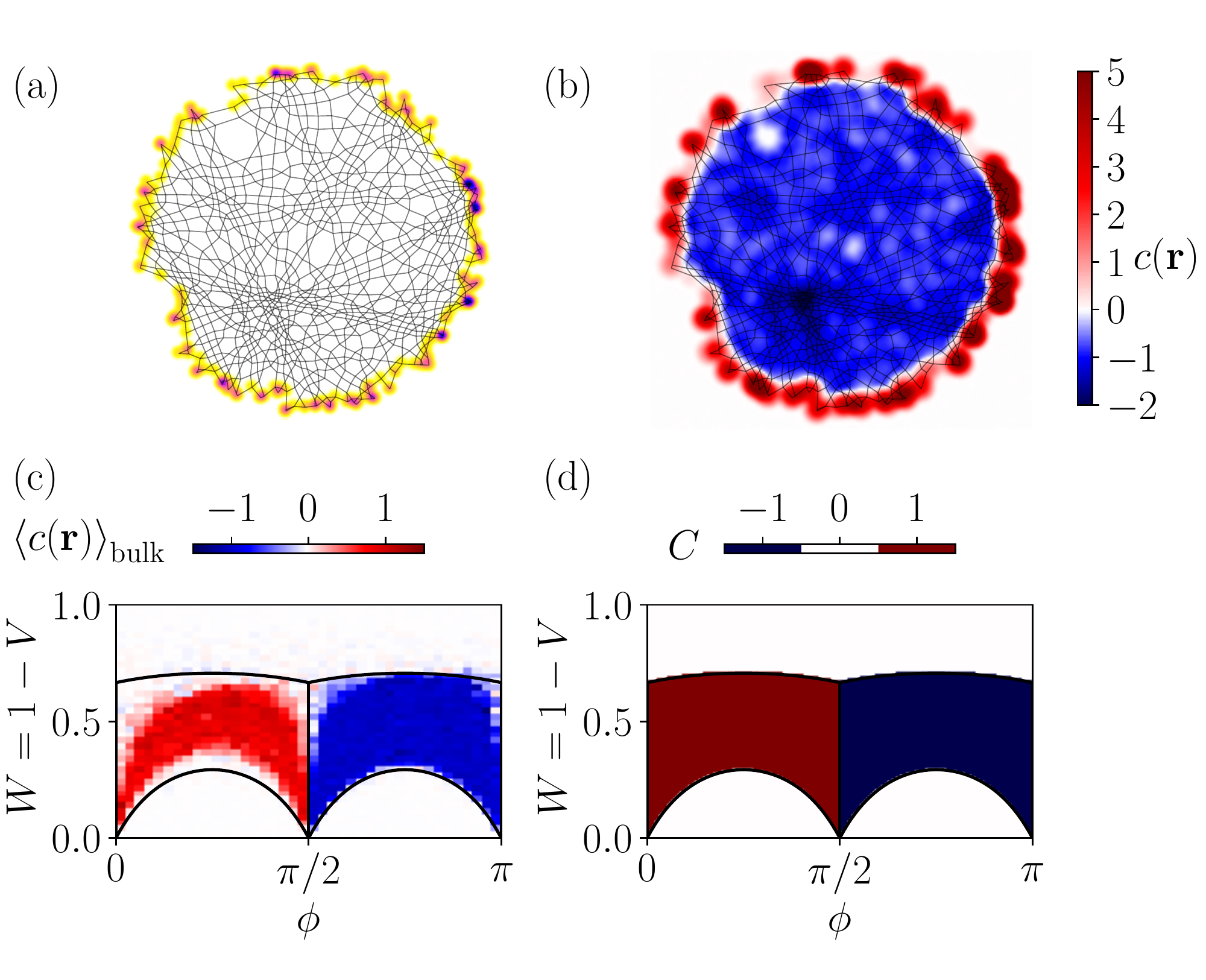}
    \caption{(a) Spatial distribution of the mid-gap edge state at half filling with $V = W$ and $\phi=2\pi/3$.
    (b) Spatial distribution of the local Chern marker density at half filling with $V = W$ and $\phi=2\pi/3$.
    (c) Phase diagram at half filling as a function of $\phi$ and $W=1-V$, the colors show the local Chern marker density averaged over the central region of the system.
    Solid lines show the phase boundaries obtained with the resolvent method.
    (d) Analytical phase diagram obtained with the resolvent method.}
    \label{fig:fourfold_chern}
\end{figure}

}

\begin{thebibliography}{56}%
\makeatletter
\providecommand \@ifxundefined [1]{%
 \@ifx{#1\undefined}
}%
\providecommand \@ifnum [1]{%
 \ifnum #1\expandafter \@firstoftwo
 \else \expandafter \@secondoftwo
 \fi
}%
\providecommand \@ifx [1]{%
 \ifx #1\expandafter \@firstoftwo
 \else \expandafter \@secondoftwo
 \fi
}%
\providecommand \natexlab [1]{#1}%
\providecommand \enquote  [1]{``#1''}%
\providecommand \bibnamefont  [1]{#1}%
\providecommand \bibfnamefont [1]{#1}%
\providecommand \citenamefont [1]{#1}%
\providecommand \href@noop [0]{\@secondoftwo}%
\providecommand \href [0]{\begingroup \@sanitize@url \@href}%
\providecommand \@href[1]{\@@startlink{#1}\@@href}%
\providecommand \@@href[1]{\endgroup#1\@@endlink}%
\providecommand \@sanitize@url [0]{\catcode `\\12\catcode `\$12\catcode
  `\&12\catcode `\#12\catcode `\^12\catcode `\_12\catcode `\%12\relax}%
\providecommand \@@startlink[1]{}%
\providecommand \@@endlink[0]{}%
\providecommand \url  [0]{\begingroup\@sanitize@url \@url }%
\providecommand \@url [1]{\endgroup\@href {#1}{\urlprefix }}%
\providecommand \urlprefix  [0]{URL }%
\providecommand \Eprint [0]{\href }%
\providecommand \doibase [0]{http://dx.doi.org/}%
\providecommand \selectlanguage [0]{\@gobble}%
\providecommand \bibinfo  [0]{\@secondoftwo}%
\providecommand \bibfield  [0]{\@secondoftwo}%
\providecommand \translation [1]{[#1]}%
\providecommand \BibitemOpen [0]{}%
\providecommand \bibitemStop [0]{}%
\providecommand \bibitemNoStop [0]{.\EOS\space}%
\providecommand \EOS [0]{\spacefactor3000\relax}%
\providecommand \BibitemShut  [1]{\csname bibitem#1\endcsname}%
\let\auto@bib@innerbib\@empty
\bibitem [{\citenamefont {Vergniory}\ \emph {et~al.}(2019)\citenamefont
  {Vergniory}, \citenamefont {Elcoro}, \citenamefont {Felser}, \citenamefont
  {Regnault}, \citenamefont {Bernevig},\ and\ \citenamefont
  {Wang}}]{Vergniory:2019ub}%
  \BibitemOpen
  \bibfield  {author} {\bibinfo {author} {\bibfnamefont {M~G}\ \bibnamefont
  {Vergniory}}, \bibinfo {author} {\bibfnamefont {L}~\bibnamefont {Elcoro}},
  \bibinfo {author} {\bibfnamefont {Claudia}\ \bibnamefont {Felser}}, \bibinfo
  {author} {\bibfnamefont {Nicolas}\ \bibnamefont {Regnault}}, \bibinfo
  {author} {\bibfnamefont {B~Andrei}\ \bibnamefont {Bernevig}}, \ and\ \bibinfo
  {author} {\bibfnamefont {Zhijun}\ \bibnamefont {Wang}},\ }\bibfield  {title}
  {\enquote {\bibinfo {title} {{A complete catalogue of high-quality
  topological materials}},}\ }\href {\doibase 10.1038/s41586-019-0954-4}
  {\bibfield  {journal} {\bibinfo  {journal} {Nature}\ }\textbf {\bibinfo
  {volume} {566}},\ \bibinfo {pages} {480--485} (\bibinfo {year}
  {2019})}\BibitemShut {NoStop}%
\bibitem [{\citenamefont {Zhang}\ \emph {et~al.}(2019)\citenamefont {Zhang},
  \citenamefont {Jiang}, \citenamefont {Song}, \citenamefont {Huang},
  \citenamefont {He}, \citenamefont {Fang}, \citenamefont {Weng},\ and\
  \citenamefont {Fang}}]{Zhang:2019tp}%
  \BibitemOpen
  \bibfield  {author} {\bibinfo {author} {\bibfnamefont {Tiantian}\
  \bibnamefont {Zhang}}, \bibinfo {author} {\bibfnamefont {Yi}~\bibnamefont
  {Jiang}}, \bibinfo {author} {\bibfnamefont {Zhida}\ \bibnamefont {Song}},
  \bibinfo {author} {\bibfnamefont {He}~\bibnamefont {Huang}}, \bibinfo
  {author} {\bibfnamefont {Yuqing}\ \bibnamefont {He}}, \bibinfo {author}
  {\bibfnamefont {Zhong}\ \bibnamefont {Fang}}, \bibinfo {author}
  {\bibfnamefont {Hongming}\ \bibnamefont {Weng}}, \ and\ \bibinfo {author}
  {\bibfnamefont {Chen}\ \bibnamefont {Fang}},\ }\bibfield  {title} {\enquote
  {\bibinfo {title} {{Catalogue of topological electronic materials}},}\ }\href
  {\doibase 10.1038/s41586-019-0944-6} {\bibfield  {journal} {\bibinfo
  {journal} {Nature}\ }\textbf {\bibinfo {volume} {566}},\ \bibinfo {pages}
  {475--479} (\bibinfo {year} {2019})}\BibitemShut {NoStop}%
\bibitem [{\citenamefont {Tang}\ \emph {et~al.}(2019)\citenamefont {Tang},
  \citenamefont {Po}, \citenamefont {Vishwanath},\ and\ \citenamefont
  {Wan}}]{Tang18}%
  \BibitemOpen
  \bibfield  {author} {\bibinfo {author} {\bibfnamefont {Feng}\ \bibnamefont
  {Tang}}, \bibinfo {author} {\bibfnamefont {Hoi~Chun}\ \bibnamefont {Po}},
  \bibinfo {author} {\bibfnamefont {Ashvin}\ \bibnamefont {Vishwanath}}, \ and\
  \bibinfo {author} {\bibfnamefont {Xiangang}\ \bibnamefont {Wan}},\ }\bibfield
   {title} {\enquote {\bibinfo {title} {{Comprehensive search for topological
  materials using symmetry indicators}},}\ }\href {\doibase
  10.1038/s41586-019-0937-5} {\bibfield  {journal} {\bibinfo  {journal}
  {Nature}\ }\textbf {\bibinfo {volume} {566}},\ \bibinfo {pages} {486--489}
  (\bibinfo {year} {2019})}\BibitemShut {NoStop}%
\bibitem [{\citenamefont {Corbae}\ \emph {et~al.}(2019)\citenamefont {Corbae},
  \citenamefont {Ciocys}, \citenamefont {Varjas}, \citenamefont {Zeltmann},
  \citenamefont {Stansbury}, \citenamefont {Molina-Ruiz}, \citenamefont {Chen},
  \citenamefont {Wang}, \citenamefont {Minor}, \citenamefont {Grushin},
  \citenamefont {Lanzara},\ and\ \citenamefont {Hellman}}]{Corbae:2019tg}%
  \BibitemOpen
  \bibfield  {author} {\bibinfo {author} {\bibfnamefont {Paul}\ \bibnamefont
  {Corbae}}, \bibinfo {author} {\bibfnamefont {Samuel}\ \bibnamefont {Ciocys}},
  \bibinfo {author} {\bibfnamefont {D{\'a}niel}\ \bibnamefont {Varjas}},
  \bibinfo {author} {\bibfnamefont {Steven}\ \bibnamefont {Zeltmann}}, \bibinfo
  {author} {\bibfnamefont {Conrad~H}\ \bibnamefont {Stansbury}}, \bibinfo
  {author} {\bibfnamefont {Manel}\ \bibnamefont {Molina-Ruiz}}, \bibinfo
  {author} {\bibfnamefont {Zhanghui}\ \bibnamefont {Chen}}, \bibinfo {author}
  {\bibfnamefont {Lin-Wang}\ \bibnamefont {Wang}}, \bibinfo {author}
  {\bibfnamefont {Andrew~M}\ \bibnamefont {Minor}}, \bibinfo {author}
  {\bibfnamefont {Adolfo~G}\ \bibnamefont {Grushin}}, \bibinfo {author}
  {\bibfnamefont {Alessandra}\ \bibnamefont {Lanzara}}, \ and\ \bibinfo
  {author} {\bibfnamefont {Frances}\ \bibnamefont {Hellman}},\ }\bibfield
  {title} {\enquote {\bibinfo {title} {{Evidence for topological surface states
  in amorphous Bi$_{2}$Se$_{3}$}},}\ }\href@noop {} {\bibfield  {journal}
  {\bibinfo  {journal} {arXiv.org}\ } (\bibinfo {year} {2019})},\ \Eprint
  {http://arxiv.org/abs/1910.13412v1} {1910.13412v1} \BibitemShut {NoStop}%
\bibitem [{\citenamefont {DC}\ \emph {et~al.}(2018)\citenamefont {DC},
  \citenamefont {Grassi}, \citenamefont {Chen}, \citenamefont {Jamali},
  \citenamefont {Reifsnyder~Hickey}, \citenamefont {Zhang}, \citenamefont
  {Zhao}, \citenamefont {Li}, \citenamefont {Quarterman}, \citenamefont {Lv},
  \citenamefont {Li}, \citenamefont {Manchon}, \citenamefont {Mkhoyan},
  \citenamefont {Low},\ and\ \citenamefont {Wang}}]{DC:2018uf}%
  \BibitemOpen
  \bibfield  {author} {\bibinfo {author} {\bibfnamefont {Mahendra}\
  \bibnamefont {DC}}, \bibinfo {author} {\bibfnamefont {Roberto}\ \bibnamefont
  {Grassi}}, \bibinfo {author} {\bibfnamefont {Jun-Yang}\ \bibnamefont {Chen}},
  \bibinfo {author} {\bibfnamefont {Mahdi}\ \bibnamefont {Jamali}}, \bibinfo
  {author} {\bibfnamefont {Danielle}\ \bibnamefont {Reifsnyder~Hickey}},
  \bibinfo {author} {\bibfnamefont {Delin}\ \bibnamefont {Zhang}}, \bibinfo
  {author} {\bibfnamefont {Zhengyang}\ \bibnamefont {Zhao}}, \bibinfo {author}
  {\bibfnamefont {Hongshi}\ \bibnamefont {Li}}, \bibinfo {author}
  {\bibfnamefont {P}~\bibnamefont {Quarterman}}, \bibinfo {author}
  {\bibfnamefont {Yang}\ \bibnamefont {Lv}}, \bibinfo {author} {\bibfnamefont
  {Mo}~\bibnamefont {Li}}, \bibinfo {author} {\bibfnamefont {Aurelien}\
  \bibnamefont {Manchon}}, \bibinfo {author} {\bibfnamefont {K~Andre}\
  \bibnamefont {Mkhoyan}}, \bibinfo {author} {\bibfnamefont {Tony}\
  \bibnamefont {Low}}, \ and\ \bibinfo {author} {\bibfnamefont {Jian-Ping}\
  \bibnamefont {Wang}},\ }\bibfield  {title} {\enquote {\bibinfo {title}
  {{Room-temperature high spin{\textendash}orbit torque due to quantum
  confinement in sputtered BixSe(1{\textendash}x) films}},}\ }\href
  {https://doi.org/10.1038/s41563-018-0136-z} {\bibfield  {journal} {\bibinfo
  {journal} {Nature materials}\ }\textbf {\bibinfo {volume} {17}},\ \bibinfo
  {pages} {800--807} (\bibinfo {year} {2018})}\BibitemShut {NoStop}%
\bibitem [{\citenamefont {{Sahu}}\ \emph {et~al.}(2019)\citenamefont {{Sahu}},
  \citenamefont {{Chen}}, \citenamefont {{Devaux}}, \citenamefont {{Jaffres}},
  \citenamefont {{Migot}}, \citenamefont {{Dang}}, \citenamefont {{George}},
  \citenamefont {{Garcia-Barriocanal}}, \citenamefont {{Lu}},\ and\
  \citenamefont {{Wang}}}]{Sahu:wy}%
  \BibitemOpen
  \bibfield  {author} {\bibinfo {author} {\bibfnamefont {Protyush}\
  \bibnamefont {{Sahu}}}, \bibinfo {author} {\bibfnamefont {Jun-Yang}\
  \bibnamefont {{Chen}}}, \bibinfo {author} {\bibfnamefont {Xavier}\
  \bibnamefont {{Devaux}}}, \bibinfo {author} {\bibfnamefont {Henri}\
  \bibnamefont {{Jaffres}}}, \bibinfo {author} {\bibfnamefont {Sylvie}\
  \bibnamefont {{Migot}}}, \bibinfo {author} {\bibfnamefont {Huong}\
  \bibnamefont {{Dang}}}, \bibinfo {author} {\bibfnamefont {Jean-Marie}\
  \bibnamefont {{George}}}, \bibinfo {author} {\bibfnamefont {Javier}\
  \bibnamefont {{Garcia-Barriocanal}}}, \bibinfo {author} {\bibfnamefont
  {Yuan}\ \bibnamefont {{Lu}}}, \ and\ \bibinfo {author} {\bibfnamefont
  {Jian-Ping}\ \bibnamefont {{Wang}}},\ }\bibfield  {title} {\enquote {\bibinfo
  {title} {{Room temperature high charge to spin conversion in amorphous
  topological insulator}},}\ }\href@noop {} {\bibfield  {journal} {\bibinfo
  {journal} {arXiv e-prints}\ ,\ \bibinfo {eid} {arXiv:1911.03323}} (\bibinfo
  {year} {2019})},\ \Eprint {http://arxiv.org/abs/1911.03323} {arXiv:1911.03323
  [cond-mat.mtrl-sci]} \BibitemShut {NoStop}%
\bibitem [{\citenamefont {Mitchell}\ \emph {et~al.}(2018)\citenamefont
  {Mitchell}, \citenamefont {Nash}, \citenamefont {Hexner}, \citenamefont
  {Turner},\ and\ \citenamefont {Irvine}}]{Mitchell2018}%
  \BibitemOpen
  \bibfield  {author} {\bibinfo {author} {\bibfnamefont {Noah~P.}\ \bibnamefont
  {Mitchell}}, \bibinfo {author} {\bibfnamefont {Lisa~M.}\ \bibnamefont
  {Nash}}, \bibinfo {author} {\bibfnamefont {Daniel}\ \bibnamefont {Hexner}},
  \bibinfo {author} {\bibfnamefont {Ari~M.}\ \bibnamefont {Turner}}, \ and\
  \bibinfo {author} {\bibfnamefont {William T.~M.}\ \bibnamefont {Irvine}},\
  }\bibfield  {title} {\enquote {\bibinfo {title} {Amorphous topological
  insulators constructed from random point sets},}\ }\href {\doibase
  10.1038/s41567-017-0024-5} {\bibfield  {journal} {\bibinfo  {journal} {Nature
  Physics}\ }\textbf {\bibinfo {volume} {14}} (\bibinfo {year} {2018}),\
  10.1038/s41567-017-0024-5}\BibitemShut {NoStop}%
\bibitem [{\citenamefont {Chalker}\ and\ \citenamefont
  {Coddington}(1988)}]{Chalker_1988}%
  \BibitemOpen
  \bibfield  {author} {\bibinfo {author} {\bibfnamefont {J~T}\ \bibnamefont
  {Chalker}}\ and\ \bibinfo {author} {\bibfnamefont {P~D}\ \bibnamefont
  {Coddington}},\ }\bibfield  {title} {\enquote {\bibinfo {title} {Percolation,
  quantum tunnelling and the integer hall effect},}\ }\href {\doibase
  10.1088/0022-3719/21/14/008} {\bibfield  {journal} {\bibinfo  {journal}
  {Journal of Physics C: Solid State Physics}\ }\textbf {\bibinfo {volume}
  {21}},\ \bibinfo {pages} {2665--2679} (\bibinfo {year} {1988})}\BibitemShut
  {NoStop}%
\bibitem [{\citenamefont {Wei}\ \emph {et~al.}(1986)\citenamefont {Wei},
  \citenamefont {Tsui},\ and\ \citenamefont {Pruisken}}]{Wei88}%
  \BibitemOpen
  \bibfield  {author} {\bibinfo {author} {\bibfnamefont {H.~P.}\ \bibnamefont
  {Wei}}, \bibinfo {author} {\bibfnamefont {D.~C.}\ \bibnamefont {Tsui}}, \
  and\ \bibinfo {author} {\bibfnamefont {A.~M.~M.}\ \bibnamefont {Pruisken}},\
  }\bibfield  {title} {\enquote {\bibinfo {title} {Localization and scaling in
  the quantum hall regime},}\ }\href {\doibase 10.1103/PhysRevB.33.1488}
  {\bibfield  {journal} {\bibinfo  {journal} {Phys. Rev. B}\ }\textbf {\bibinfo
  {volume} {33}},\ \bibinfo {pages} {1488--1491} (\bibinfo {year}
  {1986})}\BibitemShut {NoStop}%
\bibitem [{\citenamefont {Huckestein}(1995)}]{Huckestein95}%
  \BibitemOpen
  \bibfield  {author} {\bibinfo {author} {\bibfnamefont {Bodo}\ \bibnamefont
  {Huckestein}},\ }\bibfield  {title} {\enquote {\bibinfo {title} {Scaling
  theory of the integer quantum hall effect},}\ }\href {\doibase
  10.1103/RevModPhys.67.357} {\bibfield  {journal} {\bibinfo  {journal} {Rev.
  Mod. Phys.}\ }\textbf {\bibinfo {volume} {67}},\ \bibinfo {pages} {357--396}
  (\bibinfo {year} {1995})}\BibitemShut {NoStop}%
\bibitem [{\citenamefont {Agarwala}\ and\ \citenamefont
  {Shenoy}(2017)}]{Agarwala:2017jv}%
  \BibitemOpen
  \bibfield  {author} {\bibinfo {author} {\bibfnamefont {Adhip}\ \bibnamefont
  {Agarwala}}\ and\ \bibinfo {author} {\bibfnamefont {Vijay~B}\ \bibnamefont
  {Shenoy}},\ }\bibfield  {title} {{\selectlanguage {English}\enquote {\bibinfo
  {title} {{Topological Insulators in Amorphous Systems}},}\ }}\href {\doibase
  10.1103/PhysRevLett.118.236402} {\bibfield  {journal} {\bibinfo  {journal}
  {Phys. Rev. Lett.}\ }\textbf {\bibinfo {volume} {118}},\ \bibinfo {pages}
  {236402} (\bibinfo {year} {2017})}\BibitemShut {NoStop}%
\bibitem [{\citenamefont {Xiao}\ and\ \citenamefont {Fan}(2017)}]{Xiao17}%
  \BibitemOpen
  \bibfield  {author} {\bibinfo {author} {\bibfnamefont {Meng}\ \bibnamefont
  {Xiao}}\ and\ \bibinfo {author} {\bibfnamefont {Shanhui}\ \bibnamefont
  {Fan}},\ }\bibfield  {title} {\enquote {\bibinfo {title} {Photonic chern
  insulator through homogenization of an array of particles},}\ }\href
  {\doibase 10.1103/PhysRevB.96.100202} {\bibfield  {journal} {\bibinfo
  {journal} {Phys. Rev. B}\ }\textbf {\bibinfo {volume} {96}},\ \bibinfo
  {pages} {100202} (\bibinfo {year} {2017})}\BibitemShut {NoStop}%
\bibitem [{\citenamefont {P{\"o}yh{\"o}nen}\ \emph {et~al.}(2018)\citenamefont
  {P{\"o}yh{\"o}nen}, \citenamefont {Sahlberg}, \citenamefont {Weststr{\"o}m},\
  and\ \citenamefont {Ojanen}}]{Poyhonen2017}%
  \BibitemOpen
  \bibfield  {author} {\bibinfo {author} {\bibfnamefont {Kim}\ \bibnamefont
  {P{\"o}yh{\"o}nen}}, \bibinfo {author} {\bibfnamefont {Isac}\ \bibnamefont
  {Sahlberg}}, \bibinfo {author} {\bibfnamefont {Alex}\ \bibnamefont
  {Weststr{\"o}m}}, \ and\ \bibinfo {author} {\bibfnamefont {Teemu}\
  \bibnamefont {Ojanen}},\ }\bibfield  {title} {\enquote {\bibinfo {title}
  {{Amorphous topological superconductivity in a Shiba glass}},}\ }\href
  {\doibase 10.1038/s41467-018-04532-x} {\bibfield  {journal} {\bibinfo
  {journal} {Nature Communications}\ }\textbf {\bibinfo {volume} {9}},\
  \bibinfo {pages} {2103} (\bibinfo {year} {2018})}\BibitemShut {NoStop}%
\bibitem [{\citenamefont {Bourne}\ and\ \citenamefont
  {Prodan}(2018)}]{Bourne:2018jr}%
  \BibitemOpen
  \bibfield  {author} {\bibinfo {author} {\bibfnamefont {Chris}\ \bibnamefont
  {Bourne}}\ and\ \bibinfo {author} {\bibfnamefont {Emil}\ \bibnamefont
  {Prodan}},\ }\bibfield  {title} {{\selectlanguage {English}\enquote {\bibinfo
  {title} {{Non-commutative Chern numbers for generic aperiodic discrete
  systems}},}\ }}\href {\doibase 10.1088/1751-8121/aac093} {\bibfield
  {journal} {\bibinfo  {journal} {Journal of Physics A: Mathematical and
  Theoretical}\ }\textbf {\bibinfo {volume} {51}},\ \bibinfo {pages} {235202}
  (\bibinfo {year} {2018})}\BibitemShut {NoStop}%
\bibitem [{\citenamefont {Agarwala}\ \emph {et~al.}(2019)\citenamefont
  {Agarwala}, \citenamefont {Juricic},\ and\ \citenamefont
  {Roy}}]{agarwala2019higher}%
  \BibitemOpen
  \bibfield  {author} {\bibinfo {author} {\bibfnamefont {Adhip}\ \bibnamefont
  {Agarwala}}, \bibinfo {author} {\bibfnamefont {Vladimir}\ \bibnamefont
  {Juricic}}, \ and\ \bibinfo {author} {\bibfnamefont {Bitan}\ \bibnamefont
  {Roy}},\ }\href@noop {} {\enquote {\bibinfo {title} {Higher order topological
  insulators in amorphous solids},}\ } (\bibinfo {year} {2019}),\ \Eprint
  {http://arxiv.org/abs/1902.00507} {arXiv:1902.00507 [cond-mat.mes-hall]}
  \BibitemShut {NoStop}%
\bibitem [{\citenamefont {Yang}\ \emph {et~al.}(2019)\citenamefont {Yang},
  \citenamefont {Qin}, \citenamefont {Deng}, \citenamefont {Duan},\ and\
  \citenamefont {Xu}}]{Yang19}%
  \BibitemOpen
  \bibfield  {author} {\bibinfo {author} {\bibfnamefont {Yan-Bin}\ \bibnamefont
  {Yang}}, \bibinfo {author} {\bibfnamefont {Tao}\ \bibnamefont {Qin}},
  \bibinfo {author} {\bibfnamefont {Dong-Ling}\ \bibnamefont {Deng}}, \bibinfo
  {author} {\bibfnamefont {L.-M.}\ \bibnamefont {Duan}}, \ and\ \bibinfo
  {author} {\bibfnamefont {Yong}\ \bibnamefont {Xu}},\ }\bibfield  {title}
  {\enquote {\bibinfo {title} {Topological amorphous metals},}\ }\href
  {\doibase 10.1103/PhysRevLett.123.076401} {\bibfield  {journal} {\bibinfo
  {journal} {Phys. Rev. Lett.}\ }\textbf {\bibinfo {volume} {123}},\ \bibinfo
  {pages} {076401} (\bibinfo {year} {2019})}\BibitemShut {NoStop}%
\bibitem [{\citenamefont {Costa}\ \emph {et~al.}(2019)\citenamefont {Costa},
  \citenamefont {Schleder}, \citenamefont {Nardelli}, \citenamefont
  {Lewenkopf},\ and\ \citenamefont {Fazzio}}]{Costa:2019kc}%
  \BibitemOpen
  \bibfield  {author} {\bibinfo {author} {\bibfnamefont {Marcio}\ \bibnamefont
  {Costa}}, \bibinfo {author} {\bibfnamefont {Gabriel~R}\ \bibnamefont
  {Schleder}}, \bibinfo {author} {\bibfnamefont {Marco~Buongiorno}\
  \bibnamefont {Nardelli}}, \bibinfo {author} {\bibfnamefont {Caio}\
  \bibnamefont {Lewenkopf}}, \ and\ \bibinfo {author} {\bibfnamefont
  {Adalberto}\ \bibnamefont {Fazzio}},\ }\bibfield  {title} {{\selectlanguage
  {English}\enquote {\bibinfo {title} {{Toward Realistic Amorphous Topological
  Insulators}},}\ }}\href {\doibase 10.1021/acs.nanolett.9b03881} {\bibfield
  {journal} {\bibinfo  {journal} {Nano Letters}\ ,\ \bibinfo {pages}
  {acs.nanolett.9b03881--8946}} (\bibinfo {year} {2019})},\ \Eprint
  {http://arxiv.org/abs/1911.08215} {1911.08215} \BibitemShut {NoStop}%
\bibitem [{\citenamefont {Mukati}\ \emph {et~al.}(2020)\citenamefont {Mukati},
  \citenamefont {Agarwala},\ and\ \citenamefont {Bhattacharjee}}]{Mukati20}%
  \BibitemOpen
  \bibfield  {author} {\bibinfo {author} {\bibfnamefont {Prateek}\ \bibnamefont
  {Mukati}}, \bibinfo {author} {\bibfnamefont {Adhip}\ \bibnamefont
  {Agarwala}}, \ and\ \bibinfo {author} {\bibfnamefont {Subhro}\ \bibnamefont
  {Bhattacharjee}},\ }\bibfield  {title} {\enquote {\bibinfo {title}
  {Topological and conventional phases of a three-dimensional electronic
  glass},}\ }\href {\doibase 10.1103/PhysRevB.101.035142} {\bibfield  {journal}
  {\bibinfo  {journal} {Phys. Rev. B}\ }\textbf {\bibinfo {volume} {101}},\
  \bibinfo {pages} {035142} (\bibinfo {year} {2020})}\BibitemShut {NoStop}%
\bibitem [{\citenamefont {Sahlberg}\ \emph {et~al.}(2020)\citenamefont
  {Sahlberg}, \citenamefont {Weststr\"om}, \citenamefont {P\"oyh\"onen},\ and\
  \citenamefont {Ojanen}}]{Sahlberg:2019uo}%
  \BibitemOpen
  \bibfield  {author} {\bibinfo {author} {\bibfnamefont {Isac}\ \bibnamefont
  {Sahlberg}}, \bibinfo {author} {\bibfnamefont {Alex}\ \bibnamefont
  {Weststr\"om}}, \bibinfo {author} {\bibfnamefont {Kim}\ \bibnamefont
  {P\"oyh\"onen}}, \ and\ \bibinfo {author} {\bibfnamefont {Teemu}\
  \bibnamefont {Ojanen}},\ }\bibfield  {title} {\enquote {\bibinfo {title}
  {Topological phase transitions in glassy quantum matter},}\ }\href {\doibase
  10.1103/PhysRevResearch.2.013053} {\bibfield  {journal} {\bibinfo  {journal}
  {Phys. Rev. Research}\ }\textbf {\bibinfo {volume} {2}},\ \bibinfo {pages}
  {013053} (\bibinfo {year} {2020})}\BibitemShut {NoStop}%
\bibitem [{\citenamefont {{Ivaki}}\ \emph {et~al.}(2020)\citenamefont
  {{Ivaki}}, \citenamefont {{Sahlberg}},\ and\ \citenamefont
  {{Ojanen}}}]{Ivaki2020}%
  \BibitemOpen
  \bibfield  {author} {\bibinfo {author} {\bibfnamefont {Moein~N.}\
  \bibnamefont {{Ivaki}}}, \bibinfo {author} {\bibfnamefont {Isac}\
  \bibnamefont {{Sahlberg}}}, \ and\ \bibinfo {author} {\bibfnamefont {Teemu}\
  \bibnamefont {{Ojanen}}},\ }\bibfield  {title} {\enquote {\bibinfo {title}
  {{Criticality in amorphous topological matter -- beyond the universal scaling
  paradigm}},}\ }\href@noop {} {\bibfield  {journal} {\bibinfo  {journal}
  {arXiv e-prints}\ ,\ \bibinfo {eid} {arXiv:2006.05886}} (\bibinfo {year}
  {2020})},\ \Eprint {http://arxiv.org/abs/2006.05886} {arXiv:2006.05886
  [cond-mat.mes-hall]} \BibitemShut {NoStop}%
\bibitem [{\citenamefont {Kruthoff}\ \emph {et~al.}(2017)\citenamefont
  {Kruthoff}, \citenamefont {de~Boer}, \citenamefont {van Wezel}, \citenamefont
  {Kane},\ and\ \citenamefont {Slager}}]{Kruthoff17}%
  \BibitemOpen
  \bibfield  {author} {\bibinfo {author} {\bibfnamefont {Jorrit}\ \bibnamefont
  {Kruthoff}}, \bibinfo {author} {\bibfnamefont {Jan}\ \bibnamefont {de~Boer}},
  \bibinfo {author} {\bibfnamefont {Jasper}\ \bibnamefont {van Wezel}},
  \bibinfo {author} {\bibfnamefont {Charles~L.}\ \bibnamefont {Kane}}, \ and\
  \bibinfo {author} {\bibfnamefont {Robert-Jan}\ \bibnamefont {Slager}},\
  }\bibfield  {title} {\enquote {\bibinfo {title} {Topological classification
  of crystalline insulators through band structure combinatorics},}\ }\href
  {\doibase 10.1103/PhysRevX.7.041069} {\bibfield  {journal} {\bibinfo
  {journal} {Phys. Rev. X}\ }\textbf {\bibinfo {volume} {7}},\ \bibinfo {pages}
  {041069} (\bibinfo {year} {2017})}\BibitemShut {NoStop}%
\bibitem [{\citenamefont {Po}\ \emph {et~al.}(2017)\citenamefont {Po},
  \citenamefont {Vishwanath},\ and\ \citenamefont {Watanabe}}]{Po:2017ci}%
  \BibitemOpen
  \bibfield  {author} {\bibinfo {author} {\bibfnamefont {Hoi~Chun}\
  \bibnamefont {Po}}, \bibinfo {author} {\bibfnamefont {Ashvin}\ \bibnamefont
  {Vishwanath}}, \ and\ \bibinfo {author} {\bibfnamefont {Haruki}\ \bibnamefont
  {Watanabe}},\ }\bibfield  {title} {{\selectlanguage {English}\enquote
  {\bibinfo {title} {{Symmetry-based indicators of band topology in the 230
  space groups}},}\ }}\href {\doibase 10.1038/s41467-017-00133-2} {\bibfield
  {journal} {\bibinfo  {journal} {Nature Communications}\ }\textbf {\bibinfo
  {volume} {8}},\ \bibinfo {pages} {1--9} (\bibinfo {year} {2017})}\BibitemShut
  {NoStop}%
\bibitem [{\citenamefont {Bradlyn}\ \emph {et~al.}(2017)\citenamefont
  {Bradlyn}, \citenamefont {Elcoro}, \citenamefont {Cano}, \citenamefont
  {Vergniory}, \citenamefont {Wang}, \citenamefont {Felser}, \citenamefont
  {Aroyo},\ and\ \citenamefont {Bernevig}}]{Bradlyn2017}%
  \BibitemOpen
  \bibfield  {author} {\bibinfo {author} {\bibfnamefont {Barry}\ \bibnamefont
  {Bradlyn}}, \bibinfo {author} {\bibfnamefont {L}~\bibnamefont {Elcoro}},
  \bibinfo {author} {\bibfnamefont {Jennifer}\ \bibnamefont {Cano}}, \bibinfo
  {author} {\bibfnamefont {M~G}\ \bibnamefont {Vergniory}}, \bibinfo {author}
  {\bibfnamefont {Zhijun}\ \bibnamefont {Wang}}, \bibinfo {author}
  {\bibfnamefont {C}~\bibnamefont {Felser}}, \bibinfo {author} {\bibfnamefont
  {M~I}\ \bibnamefont {Aroyo}}, \ and\ \bibinfo {author} {\bibfnamefont
  {B~Andrei}\ \bibnamefont {Bernevig}},\ }\bibfield  {title} {{\selectlanguage
  {English}\enquote {\bibinfo {title} {{Topological quantum chemistry}},}\
  }}\href {\doibase 10.1038/nature23268} {\bibfield  {journal} {\bibinfo
  {journal} {Nature Publishing Group}\ }\textbf {\bibinfo {volume} {547}},\
  \bibinfo {pages} {298--305} (\bibinfo {year} {2017})}\BibitemShut {NoStop}%
\bibitem [{\citenamefont {Song}\ \emph {et~al.}(2018)\citenamefont {Song},
  \citenamefont {Zhang}, \citenamefont {Fang},\ and\ \citenamefont
  {Fang}}]{Song:2018cj}%
  \BibitemOpen
  \bibfield  {author} {\bibinfo {author} {\bibfnamefont {Zhida}\ \bibnamefont
  {Song}}, \bibinfo {author} {\bibfnamefont {Tiantian}\ \bibnamefont {Zhang}},
  \bibinfo {author} {\bibfnamefont {Zhong}\ \bibnamefont {Fang}}, \ and\
  \bibinfo {author} {\bibfnamefont {Chen}\ \bibnamefont {Fang}},\ }\bibfield
  {title} {{\selectlanguage {English}\enquote {\bibinfo {title} {{Quantitative
  mappings between symmetry and topology in solids}},}\ }}\href {\doibase
  10.1038/s41467-018-06010-w} {\bibfield  {journal} {\bibinfo  {journal}
  {Nature Communications}\ }\textbf {\bibinfo {volume} {9}},\ \bibinfo {pages}
  {3530} (\bibinfo {year} {2018})}\BibitemShut {NoStop}%
\bibitem [{\citenamefont {Zallen}(1998)}]{Zallen}%
  \BibitemOpen
  \bibfield  {author} {\bibinfo {author} {\bibfnamefont {Richard}\ \bibnamefont
  {Zallen}},\ }\href@noop {} {\emph {\bibinfo {title} {The Physics of Amorphous
  Solids}}}\ (\bibinfo  {publisher} {Wiley},\ \bibinfo {year}
  {1998})\BibitemShut {NoStop}%
\bibitem [{\citenamefont {Weaire}(1971)}]{Weaire:1971in}%
  \BibitemOpen
  \bibfield  {author} {\bibinfo {author} {\bibfnamefont {D.}~\bibnamefont
  {Weaire}},\ }\bibfield  {title} {{\selectlanguage {English}\enquote {\bibinfo
  {title} {{Existence of a Gap in the Electronic Density of States of a
  Tetrahedrally Bonded Solid of Arbitrary Structure }},}\ }}\href {\doibase
  10.1103/PhysRevLett.26.1541} {\bibfield  {journal} {\bibinfo  {journal}
  {Physical Review Letters}\ }\textbf {\bibinfo {volume} {26}},\ \bibinfo
  {pages} {1541--1543} (\bibinfo {year} {1971})}\BibitemShut {NoStop}%
\bibitem [{\citenamefont {Toh}\ \emph {et~al.}(2020)\citenamefont {Toh},
  \citenamefont {Zhang}, \citenamefont {Lin}, \citenamefont {Mayorov},
  \citenamefont {Wang}, \citenamefont {Orofeo}, \citenamefont {Ferry},
  \citenamefont {Andersen}, \citenamefont {Kakenov}, \citenamefont {Guo},
  \citenamefont {Abidi}, \citenamefont {Sims}, \citenamefont {Suenaga},
  \citenamefont {Pantelides},\ and\ \citenamefont {{\"O}zyilmaz}}]{Toh:2020dy}%
  \BibitemOpen
  \bibfield  {author} {\bibinfo {author} {\bibfnamefont {Chee-Tat}\
  \bibnamefont {Toh}}, \bibinfo {author} {\bibfnamefont {Hongji}\ \bibnamefont
  {Zhang}}, \bibinfo {author} {\bibfnamefont {Junhao}\ \bibnamefont {Lin}},
  \bibinfo {author} {\bibfnamefont {Alexander~S}\ \bibnamefont {Mayorov}},
  \bibinfo {author} {\bibfnamefont {Yun-Peng}\ \bibnamefont {Wang}}, \bibinfo
  {author} {\bibfnamefont {Carlo~M}\ \bibnamefont {Orofeo}}, \bibinfo {author}
  {\bibfnamefont {Darim~Badur}\ \bibnamefont {Ferry}}, \bibinfo {author}
  {\bibfnamefont {Henrik}\ \bibnamefont {Andersen}}, \bibinfo {author}
  {\bibfnamefont {Nurbek}\ \bibnamefont {Kakenov}}, \bibinfo {author}
  {\bibfnamefont {Zenglong}\ \bibnamefont {Guo}}, \bibinfo {author}
  {\bibfnamefont {Irfan~Haider}\ \bibnamefont {Abidi}}, \bibinfo {author}
  {\bibfnamefont {Hunter}\ \bibnamefont {Sims}}, \bibinfo {author}
  {\bibfnamefont {Kazu}\ \bibnamefont {Suenaga}}, \bibinfo {author}
  {\bibfnamefont {Sokrates~T}\ \bibnamefont {Pantelides}}, \ and\ \bibinfo
  {author} {\bibfnamefont {Barbaros}\ \bibnamefont {{\"O}zyilmaz}},\ }\bibfield
   {title} {{\selectlanguage {English}\enquote {\bibinfo {title} {{Synthesis
  and properties of free-standing monolayer amorphous carbon}},}\ }}\href
  {\doibase 10.1038/s41586-019-1871-2} {\bibfield  {journal} {\bibinfo
  {journal} {Nature Publishing Group}\ }\textbf {\bibinfo {volume} {577}},\
  \bibinfo {pages} {199--203} (\bibinfo {year} {2020})}\BibitemShut {NoStop}%
\bibitem [{\citenamefont {Weaire}\ and\ \citenamefont
  {Thorpe}(1971)}]{Weaire71}%
  \BibitemOpen
  \bibfield  {author} {\bibinfo {author} {\bibfnamefont {D.}~\bibnamefont
  {Weaire}}\ and\ \bibinfo {author} {\bibfnamefont {M.~F.}\ \bibnamefont
  {Thorpe}},\ }\bibfield  {title} {\enquote {\bibinfo {title} {Electronic
  properties of an amorphous solid. i. a simple tight-binding theory},}\ }\href
  {\doibase 10.1103/PhysRevB.4.2508} {\bibfield  {journal} {\bibinfo  {journal}
  {Phys. Rev. B}\ }\textbf {\bibinfo {volume} {4}},\ \bibinfo {pages}
  {2508--2520} (\bibinfo {year} {1971})}\BibitemShut {NoStop}%
\bibitem [{\citenamefont {Thorpe}\ \emph {et~al.}(1973)\citenamefont {Thorpe},
  \citenamefont {Weaire},\ and\ \citenamefont {Alben}}]{Thorpe:1973cr}%
  \BibitemOpen
  \bibfield  {author} {\bibinfo {author} {\bibfnamefont {M~F}\ \bibnamefont
  {Thorpe}}, \bibinfo {author} {\bibfnamefont {D}~\bibnamefont {Weaire}}, \
  and\ \bibinfo {author} {\bibfnamefont {R}~\bibnamefont {Alben}},\ }\bibfield
  {title} {{\selectlanguage {English}\enquote {\bibinfo {title} {{Electronic
  Properties of an Amorphous Solid. III. The Cohesive Energy and the Density of
  States}},}\ }}\href {\doibase 10.1103/PhysRevB.7.3777} {\bibfield  {journal}
  {\bibinfo  {journal} {Physical Review B}\ }\textbf {\bibinfo {volume} {7}},\
  \bibinfo {pages} {3777--3788} (\bibinfo {year} {1973})}\BibitemShut {NoStop}%
\bibitem [{\citenamefont {Bianco}\ and\ \citenamefont {Resta}(2011)}]{LCM}%
  \BibitemOpen
  \bibfield  {author} {\bibinfo {author} {\bibfnamefont {Raffaello}\
  \bibnamefont {Bianco}}\ and\ \bibinfo {author} {\bibfnamefont {Raffaele}\
  \bibnamefont {Resta}},\ }\bibfield  {title} {\enquote {\bibinfo {title}
  {Mapping topological order in coordinate space},}\ }\href {\doibase
  10.1103/PhysRevB.84.241106} {\bibfield  {journal} {\bibinfo  {journal} {Phys.
  Rev. B}\ }\textbf {\bibinfo {volume} {84}},\ \bibinfo {pages} {241106}
  (\bibinfo {year} {2011})}\BibitemShut {NoStop}%
\bibitem [{SuppMat()}]{SuppMat}%
  \BibitemOpen
  \bibinfo {note} {The supplementary material includes further information on
  the local Chern marker, circular dichroism, the properties of $z-$fold
  coordinated models, the resolvent method, symmetries of eigenstates and the
  effective Hamiltonian.}\BibitemShut {Stop}%
\bibitem [{\citenamefont {Varjas}\ \emph {et~al.}(2019)\citenamefont {Varjas},
  \citenamefont {Lau}, \citenamefont {P\"oyh\"onen}, \citenamefont {Akhmerov},
  \citenamefont {Pikulin},\ and\ \citenamefont {Fulga}}]{Varjas2019}%
  \BibitemOpen
  \bibfield  {author} {\bibinfo {author} {\bibfnamefont {D\'aniel}\
  \bibnamefont {Varjas}}, \bibinfo {author} {\bibfnamefont {Alexander}\
  \bibnamefont {Lau}}, \bibinfo {author} {\bibfnamefont {Kim}\ \bibnamefont
  {P\"oyh\"onen}}, \bibinfo {author} {\bibfnamefont {Anton~R.}\ \bibnamefont
  {Akhmerov}}, \bibinfo {author} {\bibfnamefont {Dmitry~I.}\ \bibnamefont
  {Pikulin}}, \ and\ \bibinfo {author} {\bibfnamefont {Ion~Cosma}\ \bibnamefont
  {Fulga}},\ }\bibfield  {title} {\enquote {\bibinfo {title} {Topological
  phases without crystalline counterparts},}\ }\href {\doibase
  10.1103/PhysRevLett.123.196401} {\bibfield  {journal} {\bibinfo  {journal}
  {Phys. Rev. Lett.}\ }\textbf {\bibinfo {volume} {123}},\ \bibinfo {pages}
  {196401} (\bibinfo {year} {2019})}\BibitemShut {NoStop}%
\bibitem [{\citenamefont {Schwartz}\ and\ \citenamefont
  {Ehrenreich}(1972)}]{Schwartz72}%
  \BibitemOpen
  \bibfield  {author} {\bibinfo {author} {\bibfnamefont {L.}~\bibnamefont
  {Schwartz}}\ and\ \bibinfo {author} {\bibfnamefont {H.}~\bibnamefont
  {Ehrenreich}},\ }\bibfield  {title} {\enquote {\bibinfo {title} {Comment on
  the tight-binding model for amorphous semiconductors},}\ }\href {\doibase
  10.1103/PhysRevB.6.4088} {\bibfield  {journal} {\bibinfo  {journal} {Phys.
  Rev. B}\ }\textbf {\bibinfo {volume} {6}},\ \bibinfo {pages} {4088--4090}
  (\bibinfo {year} {1972})}\BibitemShut {NoStop}%
\bibitem [{\citenamefont {Wei\ss{}e}\ \emph {et~al.}(2006)\citenamefont
  {Wei\ss{}e}, \citenamefont {Wellein}, \citenamefont {Alvermann},\ and\
  \citenamefont {Fehske}}]{KPM}%
  \BibitemOpen
  \bibfield  {author} {\bibinfo {author} {\bibfnamefont {Alexander}\
  \bibnamefont {Wei\ss{}e}}, \bibinfo {author} {\bibfnamefont {Gerhard}\
  \bibnamefont {Wellein}}, \bibinfo {author} {\bibfnamefont {Andreas}\
  \bibnamefont {Alvermann}}, \ and\ \bibinfo {author} {\bibfnamefont {Holger}\
  \bibnamefont {Fehske}},\ }\bibfield  {title} {\enquote {\bibinfo {title} {The
  kernel polynomial method},}\ }\href {\doibase 10.1103/RevModPhys.78.275}
  {\bibfield  {journal} {\bibinfo  {journal} {Rev. Mod. Phys.}\ }\textbf
  {\bibinfo {volume} {78}},\ \bibinfo {pages} {275--306} (\bibinfo {year}
  {2006})}\BibitemShut {NoStop}%
\bibitem [{\citenamefont {Bianco}(2014)}]{Bianco_thesis}%
  \BibitemOpen
  \bibfield  {author} {\bibinfo {author} {\bibfnamefont {Raffaello}\
  \bibnamefont {Bianco}},\ }\href@noop {} {\emph {\bibinfo {title} {Chern
  invariant and orbital magnetization as local quantities}}}\ (\bibinfo
  {publisher} {Università degli studi di Trieste},\ \bibinfo {year}
  {2014})\BibitemShut {NoStop}%
\bibitem [{\citenamefont {Tran}\ \emph {et~al.}(2017)\citenamefont {Tran},
  \citenamefont {Dauphin}, \citenamefont {Grushin}, \citenamefont {Zoller},\
  and\ \citenamefont {Goldman}}]{Tran2017}%
  \BibitemOpen
  \bibfield  {author} {\bibinfo {author} {\bibfnamefont {Duc~Thanh}\
  \bibnamefont {Tran}}, \bibinfo {author} {\bibfnamefont {Alexandre}\
  \bibnamefont {Dauphin}}, \bibinfo {author} {\bibfnamefont {Adolfo~G}\
  \bibnamefont {Grushin}}, \bibinfo {author} {\bibfnamefont {Peter}\
  \bibnamefont {Zoller}}, \ and\ \bibinfo {author} {\bibfnamefont {Nathan}\
  \bibnamefont {Goldman}},\ }\bibfield  {title} {{\selectlanguage
  {English}\enquote {\bibinfo {title} {{Probing topology by
  {\textquotedblleft}heating{\textquotedblright}: Quantized circular dichroism
  in ultracold atoms}},}\ }}\href {\doibase 10.1126/sciadv.1701207} {\bibfield
  {journal} {\bibinfo  {journal} {Science Advances}\ }\textbf {\bibinfo
  {volume} {3}},\ \bibinfo {pages} {e1701207} (\bibinfo {year}
  {2017})}\BibitemShut {NoStop}%
\bibitem [{\citenamefont {Pozo}\ \emph {et~al.}(2019)\citenamefont {Pozo},
  \citenamefont {Repellin},\ and\ \citenamefont {Grushin}}]{Pozo2019}%
  \BibitemOpen
  \bibfield  {author} {\bibinfo {author} {\bibfnamefont {Oscar}\ \bibnamefont
  {Pozo}}, \bibinfo {author} {\bibfnamefont {C\'ecile}\ \bibnamefont
  {Repellin}}, \ and\ \bibinfo {author} {\bibfnamefont {Adolfo~G.}\
  \bibnamefont {Grushin}},\ }\bibfield  {title} {\enquote {\bibinfo {title}
  {Quantization in chiral higher order topological insulators: Circular
  dichroism and local chern marker},}\ }\href {\doibase
  10.1103/PhysRevLett.123.247401} {\bibfield  {journal} {\bibinfo  {journal}
  {Phys. Rev. Lett.}\ }\textbf {\bibinfo {volume} {123}},\ \bibinfo {pages}
  {247401} (\bibinfo {year} {2019})}\BibitemShut {NoStop}%
\bibitem [{\citenamefont {Thouless}(1984)}]{Thouless:1984en}%
  \BibitemOpen
  \bibfield  {author} {\bibinfo {author} {\bibfnamefont {D~J}\ \bibnamefont
  {Thouless}},\ }\bibfield  {title} {{\selectlanguage {English}\enquote
  {\bibinfo {title} {{Wannier functions for magnetic sub-bands}},}\ }}\href
  {\doibase 10.1088/0022-3719/17/12/003} {\bibfield  {journal} {\bibinfo
  {journal} {Journal of Physics C: Solid State Physics}\ }\textbf {\bibinfo
  {volume} {17}},\ \bibinfo {pages} {L325--L327} (\bibinfo {year}
  {1984})}\BibitemShut {NoStop}%
\bibitem [{\citenamefont {Soluyanov}\ and\ \citenamefont
  {Vanderbilt}(2011)}]{Soluyanov2011}%
  \BibitemOpen
  \bibfield  {author} {\bibinfo {author} {\bibfnamefont {Alexey~A.}\
  \bibnamefont {Soluyanov}}\ and\ \bibinfo {author} {\bibfnamefont {David}\
  \bibnamefont {Vanderbilt}},\ }\bibfield  {title} {\enquote {\bibinfo {title}
  {Wannier representation of $\mathbb{Z}_{2}$ topological insulators},}\ }\href
  {\doibase 10.1103/PhysRevB.83.035108} {\bibfield  {journal} {\bibinfo
  {journal} {Phys. Rev. B}\ }\textbf {\bibinfo {volume} {83}},\ \bibinfo
  {pages} {035108} (\bibinfo {year} {2011})}\BibitemShut {NoStop}%
\bibitem [{\citenamefont {{Po}}(2020)}]{Po20}%
  \BibitemOpen
  \bibfield  {author} {\bibinfo {author} {\bibfnamefont {Hoi~Chun}\
  \bibnamefont {{Po}}},\ }\bibfield  {title} {\enquote {\bibinfo {title}
  {{Symmetry indicators of band topology}},}\ }\href@noop {} {\bibfield
  {journal} {\bibinfo  {journal} {arXiv e-prints}\ ,\ \bibinfo {eid}
  {arXiv:2002.09391}} (\bibinfo {year} {2020})},\ \Eprint
  {http://arxiv.org/abs/2002.09391} {arXiv:2002.09391 [cond-mat.mes-hall]}
  \BibitemShut {NoStop}%
\bibitem [{\citenamefont {Fang}\ \emph {et~al.}(2012)\citenamefont {Fang},
  \citenamefont {Gilbert},\ and\ \citenamefont {Bernevig}}]{Fang2012}%
  \BibitemOpen
  \bibfield  {author} {\bibinfo {author} {\bibfnamefont {Chen}\ \bibnamefont
  {Fang}}, \bibinfo {author} {\bibfnamefont {Matthew~J.}\ \bibnamefont
  {Gilbert}}, \ and\ \bibinfo {author} {\bibfnamefont {B.~Andrei}\ \bibnamefont
  {Bernevig}},\ }\bibfield  {title} {\enquote {\bibinfo {title} {Bulk
  topological invariants in noninteracting point group symmetric insulators},}\
  }\href {\doibase 10.1103/PhysRevB.86.115112} {\bibfield  {journal} {\bibinfo
  {journal} {Phys. Rev. B}\ }\textbf {\bibinfo {volume} {86}},\ \bibinfo
  {pages} {115112} (\bibinfo {year} {2012})}\BibitemShut {NoStop}%
\bibitem [{\citenamefont {Van~Mechelen}\ and\ \citenamefont
  {Jacob}(2018)}]{VanMechelen:2018cy}%
  \BibitemOpen
  \bibfield  {author} {\bibinfo {author} {\bibfnamefont {Todd}\ \bibnamefont
  {Van~Mechelen}}\ and\ \bibinfo {author} {\bibfnamefont {Zubin}\ \bibnamefont
  {Jacob}},\ }\bibfield  {title} {{\selectlanguage {English}\enquote {\bibinfo
  {title} {{Quantum gyroelectric effect: Photon spin-1 quantization in
  continuum topological bosonic phases}},}\ }}\href {\doibase
  10.1103/PhysRevA.98.023842} {\bibfield  {journal} {\bibinfo  {journal}
  {Physical Review A}\ }\textbf {\bibinfo {volume} {98}},\ \bibinfo {pages} {1}
  (\bibinfo {year} {2018})}\BibitemShut {NoStop}%
\bibitem [{\citenamefont {Van~Mechelen}\ and\ \citenamefont
  {Jacob}(2019)}]{VanMechelen:2019ha}%
  \BibitemOpen
  \bibfield  {author} {\bibinfo {author} {\bibfnamefont {Todd}\ \bibnamefont
  {Van~Mechelen}}\ and\ \bibinfo {author} {\bibfnamefont {Zubin}\ \bibnamefont
  {Jacob}},\ }\bibfield  {title} {{\selectlanguage {English}\enquote {\bibinfo
  {title} {{Nonlocal topological electromagnetic phases of matter}},}\ }}\href
  {\doibase 10.1103/PhysRevB.99.205146} {\bibfield  {journal} {\bibinfo
  {journal} {Physical Review B}\ }\textbf {\bibinfo {volume} {99}},\ \bibinfo
  {pages} {205146} (\bibinfo {year} {2019})}\BibitemShut {NoStop}%
\bibitem [{\citenamefont {Florescu}\ \emph {et~al.}(2009)\citenamefont
  {Florescu}, \citenamefont {Torquato},\ and\ \citenamefont
  {Steinhardt}}]{Florescu:2009ev}%
  \BibitemOpen
  \bibfield  {author} {\bibinfo {author} {\bibfnamefont {Marian}\ \bibnamefont
  {Florescu}}, \bibinfo {author} {\bibfnamefont {Salvatore}\ \bibnamefont
  {Torquato}}, \ and\ \bibinfo {author} {\bibfnamefont {Paul~J}\ \bibnamefont
  {Steinhardt}},\ }\bibfield  {title} {{\selectlanguage {English}\enquote
  {\bibinfo {title} {{Designer disordered materials with large, complete
  photonic band gaps}},}\ }}\href {\doibase 10.1073/pnas.0907744106} {\bibfield
   {journal} {\bibinfo  {journal} {Proceedings of the National Academy of
  Sciences}\ }\textbf {\bibinfo {volume} {106}},\ \bibinfo {pages}
  {20658--20663} (\bibinfo {year} {2009})}\BibitemShut {NoStop}%
\bibitem [{\citenamefont {Rechtsman}\ \emph {et~al.}(2011)\citenamefont
  {Rechtsman}, \citenamefont {Szameit}, \citenamefont {Dreisow}, \citenamefont
  {Heinrich}, \citenamefont {Keil}, \citenamefont {Nolte},\ and\ \citenamefont
  {Segev}}]{Rechtsman2011}%
  \BibitemOpen
  \bibfield  {author} {\bibinfo {author} {\bibfnamefont {Mikael}\ \bibnamefont
  {Rechtsman}}, \bibinfo {author} {\bibfnamefont {Alexander}\ \bibnamefont
  {Szameit}}, \bibinfo {author} {\bibfnamefont {Felix}\ \bibnamefont
  {Dreisow}}, \bibinfo {author} {\bibfnamefont {Matthias}\ \bibnamefont
  {Heinrich}}, \bibinfo {author} {\bibfnamefont {Robert}\ \bibnamefont {Keil}},
  \bibinfo {author} {\bibfnamefont {Stefan}\ \bibnamefont {Nolte}}, \ and\
  \bibinfo {author} {\bibfnamefont {Mordechai}\ \bibnamefont {Segev}},\
  }\bibfield  {title} {\enquote {\bibinfo {title} {Amorphous photonic lattices:
  Band gaps, effective mass, and suppressed transport},}\ }\href {\doibase
  10.1103/PhysRevLett.106.193904} {\bibfield  {journal} {\bibinfo  {journal}
  {Phys. Rev. Lett.}\ }\textbf {\bibinfo {volume} {106}},\ \bibinfo {pages}
  {193904} (\bibinfo {year} {2011})}\BibitemShut {NoStop}%
\bibitem [{\citenamefont {Groth}\ \emph {et~al.}(2014)\citenamefont {Groth},
  \citenamefont {Wimmer}, \citenamefont {Akhmerov},\ and\ \citenamefont
  {Waintal}}]{Groth_2014}%
  \BibitemOpen
  \bibfield  {author} {\bibinfo {author} {\bibfnamefont {Christoph~W}\
  \bibnamefont {Groth}}, \bibinfo {author} {\bibfnamefont {Michael}\
  \bibnamefont {Wimmer}}, \bibinfo {author} {\bibfnamefont {Anton~R}\
  \bibnamefont {Akhmerov}}, \ and\ \bibinfo {author} {\bibfnamefont {Xavier}\
  \bibnamefont {Waintal}},\ }\bibfield  {title} {\enquote {\bibinfo {title}
  {Kwant: a software package for quantum transport},}\ }\href {\doibase
  10.1088/1367-2630/16/6/063065} {\bibfield  {journal} {\bibinfo  {journal}
  {New Journal of Physics}\ }\textbf {\bibinfo {volume} {16}},\ \bibinfo
  {pages} {063065} (\bibinfo {year} {2014})}\BibitemShut {NoStop}%
\bibitem [{\citenamefont {Hunter}(2007)}]{Hunter:2007}%
  \BibitemOpen
  \bibfield  {author} {\bibinfo {author} {\bibfnamefont {J.~D.}\ \bibnamefont
  {Hunter}},\ }\bibfield  {title} {\enquote {\bibinfo {title} {Matplotlib: A 2d
  graphics environment},}\ }\href {\doibase 10.1109/MCSE.2007.55} {\bibfield
  {journal} {\bibinfo  {journal} {Computing in Science \& Engineering}\
  }\textbf {\bibinfo {volume} {9}},\ \bibinfo {pages} {90--95} (\bibinfo {year}
  {2007})}\BibitemShut {NoStop}%
\bibitem [{\citenamefont {Marsal}\ \emph {et~al.}(2020)\citenamefont {Marsal},
  \citenamefont {Varjas},\ and\ \citenamefont {Grushin}}]{zenodo}%
  \BibitemOpen
  \bibfield  {author} {\bibinfo {author} {\bibfnamefont {Quentin}\ \bibnamefont
  {Marsal}}, \bibinfo {author} {\bibfnamefont {D{\'a}niel}\ \bibnamefont
  {Varjas}}, \ and\ \bibinfo {author} {\bibfnamefont {Adolfo~G.}\ \bibnamefont
  {Grushin}},\ }\bibfield  {title} {\enquote {\bibinfo {title} {Topological
  {Weaire-Thorpe} models of amorphous matter},}\ }\href
  {https://doi.org/10.5281/zenodo.3741829} {\bibfield  {journal} {\bibinfo
  {journal} {zenodo.3741829}\ } (\bibinfo {year} {2020})}\BibitemShut {NoStop}%
\bibitem [{\citenamefont {Haldane}(2004)}]{Haldane04}%
  \BibitemOpen
  \bibfield  {author} {\bibinfo {author} {\bibfnamefont {F.~D.~M.}\
  \bibnamefont {Haldane}},\ }\bibfield  {title} {\enquote {\bibinfo {title}
  {Berry curvature on the fermi surface: Anomalous hall effect as a topological
  fermi-liquid property},}\ }\href {\doibase 10.1103/PhysRevLett.93.206602}
  {\bibfield  {journal} {\bibinfo  {journal} {Phys. Rev. Lett.}\ }\textbf
  {\bibinfo {volume} {93}},\ \bibinfo {pages} {206602} (\bibinfo {year}
  {2004})}\BibitemShut {NoStop}%
\bibitem [{\citenamefont {Tran}\ \emph {et~al.}(2015)\citenamefont {Tran},
  \citenamefont {Dauphin}, \citenamefont {Goldman},\ and\ \citenamefont
  {Gaspard}}]{Tran:2015cj}%
  \BibitemOpen
  \bibfield  {author} {\bibinfo {author} {\bibfnamefont {Duc~Thanh}\
  \bibnamefont {Tran}}, \bibinfo {author} {\bibfnamefont {Alexandre}\
  \bibnamefont {Dauphin}}, \bibinfo {author} {\bibfnamefont {Nathan}\
  \bibnamefont {Goldman}}, \ and\ \bibinfo {author} {\bibfnamefont {Pierre}\
  \bibnamefont {Gaspard}},\ }\bibfield  {title} {\enquote {\bibinfo {title}
  {{Topological Hofstadter insulators in a two-dimensional quasicrystal}},}\
  }\href {https://journals.aps.org/prb/abstract/10.1103/PhysRevB.91.085125}
  {\bibfield  {journal} {\bibinfo  {journal} {Physical Review B}\ }\textbf
  {\bibinfo {volume} {91}},\ \bibinfo {pages} {085125} (\bibinfo {year}
  {2015})}\BibitemShut {NoStop}%
\bibitem [{\citenamefont {Loring}\ and\ \citenamefont
  {Hastings}(2010)}]{Loring:2010jh}%
  \BibitemOpen
  \bibfield  {author} {\bibinfo {author} {\bibfnamefont {T~A}\ \bibnamefont
  {Loring}}\ and\ \bibinfo {author} {\bibfnamefont {M~B}\ \bibnamefont
  {Hastings}},\ }\bibfield  {title} {{\selectlanguage {English}\enquote
  {\bibinfo {title} {{Disordered topological insulators via C*-algebras}},}\
  }}\href {\doibase 10.1209/0295-5075/92/67004} {\bibfield  {journal} {\bibinfo
   {journal} {EPL (Europhysics Letters)}\ }\textbf {\bibinfo {volume} {92}},\
  \bibinfo {pages} {67004} (\bibinfo {year} {2010})}\BibitemShut {NoStop}%
\bibitem [{\citenamefont {Souza}\ and\ \citenamefont
  {Vanderbilt}(2008)}]{Souza2008}%
  \BibitemOpen
  \bibfield  {author} {\bibinfo {author} {\bibfnamefont {Ivo}\ \bibnamefont
  {Souza}}\ and\ \bibinfo {author} {\bibfnamefont {David}\ \bibnamefont
  {Vanderbilt}},\ }\bibfield  {title} {\enquote {\bibinfo {title} {{Dichroic f
  -sum rule and the orbital magnetization of crystals}},}\ }\href {\doibase
  10.1103/PhysRevB.77.054438} {\bibfield  {journal} {\bibinfo  {journal}
  {Physical Review B - Condensed Matter and Materials Physics}\ }\textbf
  {\bibinfo {volume} {77}},\ \bibinfo {pages} {1--13} (\bibinfo {year}
  {2008})}\BibitemShut {NoStop}%
\bibitem [{\citenamefont {Asteria}\ \emph {et~al.}(2019)\citenamefont
  {Asteria}, \citenamefont {Tran}, \citenamefont {Ozawa}, \citenamefont
  {Tarnowski}, \citenamefont {Rem}, \citenamefont {Fl{\"a}schner},
  \citenamefont {Sengstock}, \citenamefont {Goldman},\ and\ \citenamefont
  {Weitenberg}}]{Asteria:2019if}%
  \BibitemOpen
  \bibfield  {author} {\bibinfo {author} {\bibfnamefont {Luca}\ \bibnamefont
  {Asteria}}, \bibinfo {author} {\bibfnamefont {Duc~Thanh}\ \bibnamefont
  {Tran}}, \bibinfo {author} {\bibfnamefont {Tomoki}\ \bibnamefont {Ozawa}},
  \bibinfo {author} {\bibfnamefont {Matthias}\ \bibnamefont {Tarnowski}},
  \bibinfo {author} {\bibfnamefont {Benno~S}\ \bibnamefont {Rem}}, \bibinfo
  {author} {\bibfnamefont {Nick}\ \bibnamefont {Fl{\"a}schner}}, \bibinfo
  {author} {\bibfnamefont {Klaus}\ \bibnamefont {Sengstock}}, \bibinfo {author}
  {\bibfnamefont {Nathan}\ \bibnamefont {Goldman}}, \ and\ \bibinfo {author}
  {\bibfnamefont {Christof}\ \bibnamefont {Weitenberg}},\ }\bibfield  {title}
  {{\selectlanguage {English}\enquote {\bibinfo {title} {{Measuring quantized
  circular dichroism in ultracold topological matter}},}\ }}\href {\doibase
  10.1038/s41567-019-0417-8} {\bibfield  {journal} {\bibinfo  {journal} {Nature
  Physics}\ }\textbf {\bibinfo {volume} {49}},\ \bibinfo {pages} {1} (\bibinfo
  {year} {2019})}\BibitemShut {NoStop}%
\bibitem [{\citenamefont {Wei{\ss}e}\ \emph {et~al.}(2006)\citenamefont
  {Wei{\ss}e}, \citenamefont {Wellein}, \citenamefont {Alvermann},\ and\
  \citenamefont {Fehske}}]{weisse2006kernel}%
  \BibitemOpen
  \bibfield  {author} {\bibinfo {author} {\bibfnamefont {Alexander}\
  \bibnamefont {Wei{\ss}e}}, \bibinfo {author} {\bibfnamefont {Gerhard}\
  \bibnamefont {Wellein}}, \bibinfo {author} {\bibfnamefont {Andreas}\
  \bibnamefont {Alvermann}}, \ and\ \bibinfo {author} {\bibfnamefont {Holger}\
  \bibnamefont {Fehske}},\ }\bibfield  {title} {\enquote {\bibinfo {title} {The
  kernel polynomial method},}\ }\href
  {https://link.aps.org/doi/10.1103/RevModPhys.78.275} {\bibfield  {journal}
  {\bibinfo  {journal} {Rev. Mod. Phys.}\ }\textbf {\bibinfo {volume} {78}},\
  \bibinfo {pages} {275} (\bibinfo {year} {2006})}\BibitemShut {NoStop}%
\bibitem [{\citenamefont {{Varjas}}\ \emph {et~al.}(2020)\citenamefont
  {{Varjas}}, \citenamefont {{Fruchart}}, \citenamefont {{Akhmerov}},\ and\
  \citenamefont {{Perez-Piskunow}}}]{Varjas2019b}%
  \BibitemOpen
  \bibfield  {author} {\bibinfo {author} {\bibfnamefont {D{\'a}niel}\
  \bibnamefont {{Varjas}}}, \bibinfo {author} {\bibfnamefont {Michel}\
  \bibnamefont {{Fruchart}}}, \bibinfo {author} {\bibfnamefont {Anton~R.}\
  \bibnamefont {{Akhmerov}}}, \ and\ \bibinfo {author} {\bibfnamefont
  {Pablo~M.}\ \bibnamefont {{Perez-Piskunow}}},\ }\bibfield  {title} {\enquote
  {\bibinfo {title} {{Computation of topological phase diagram of disordered
  Pb$_{1 -x}$Sn$_{x}$Te using the kernel polynomial method}},}\ }\href
  {\doibase 10.1103/PhysRevResearch.2.013229} {\bibfield  {journal} {\bibinfo
  {journal} {Physical Review Research}\ }\textbf {\bibinfo {volume} {2}},\
  \bibinfo {eid} {013229} (\bibinfo {year} {2020})},\ \Eprint
  {http://arxiv.org/abs/1905.02215} {arXiv:1905.02215 [cond-mat.mes-hall]}
  \BibitemShut {NoStop}%
\bibitem [{\citenamefont {Miles}(1964)}]{Miles1964}%
  \BibitemOpen
  \bibfield  {author} {\bibinfo {author} {\bibfnamefont {R~E}\ \bibnamefont
  {Miles}},\ }\bibfield  {title} {\enquote {\bibinfo {title} {Random polygons
  determined by random lines in a plane},}\ }\href {\doibase
  10.1073/pnas.52.4.901} {\bibfield  {journal} {\bibinfo  {journal}
  {Proceedings of the National Academy of Sciences of the United States of
  America}\ }\textbf {\bibinfo {volume} {52}},\ \bibinfo {pages} {901--907}
  (\bibinfo {year} {1964})}\BibitemShut {NoStop}%
\end{thebibliography}
\end{document}